\documentclass[preprint,10pt]{JHEP3} 

\JHEPspecialurl{http://jhep.sissa.it/JOURNAL/JHEP3.tar.gz}

\usepackage{epsfig,multicol}

\def\beq{\begin{equation}}
\def\eeq{\end{equation}}
\def\bea{\begin{eqnarray}}
\def\eea{\end{eqnarray}}

\def\eq#1{{Eq.~(\ref{#1})}}
\def\fig#1{{Fig.~\ref{#1}}}
\newcommand{\bas}{\bar{\alpha}_S}
\newcommand{\as}{\alpha_S}

\newcommand{\Lb}{\left(}
\newcommand{\Rb}{\right)}

\parskip=\itemsep               

\newcommand{\D}{\partial}
\def\thefootnote{\fnsymbol{footnote}}

\title{\LARGE \bf  The BFKL Pomeron Calculus in the dipole approach}
\author{\Large M. Kozlov \thanks{Email: kozlov@post.tau.ac.il}\,\,,\,\,E. Levin\thanks{Email:
leving@post.tau.ac.il, levin@mail.desy.de}\,\,  and\,\, A.
Prygarin\thanks{Email:
prygarin@post.tau.ac.il} \\
Department of Particle Physics, School of Physics and Astronomy\\ Raymond and Beverly Sackler Faculty
of Exact Science\\  Tel Aviv University, Tel Aviv, 69978, Israel}

\abstract{In this paper we continue to pursue a goal of finding an effective theory for high energy
interaction
in QCD based on the colour dipole approach, for which the BFKL Pomeron Calculus gives a low energy
limit.   The  key problem, that we try to solve in this paper is
the probabilistic interpretation of the BFKL Pomeron Calculus in terms of the colourless dipoles and their interactions.
We demonstrate that the BFKL Pomeron Calculus has two equivalent descriptions :
(i)\,\,one is the generating functional  which  gives a  clear
probabilistic
interpretation of the processes of high energy scattering and  also provides  a Hamiltonian-like
description of the
system
of interacting dipoles; (ii)\,\,the second is the Langevin equation with a specific noise term which is rather complicated.
We found that at high energies this Langevin equation can be reduced to  the Langevin equation for directed percolation in the momentum space if the impact parameter is large, namely, $b>> 1/k$, where $k$ is the transverse momentum of a dipole.
Unfortunately, this simplified form of
Langevin equation is not applicable for summation of Pomeron loops, where one integrates over all
possible values of impact parameter.
We show that the BFKL Pomeron calculus with two vertices (
splitting $P\rightarrow P+P$ and merging $P+P\rightarrow P$ of Pomerons )
can be interpreted as a system of colourless dipoles with two processes: the decay of one dipole into two and
the merging of two dipoles into one dipole. However, a number of assumptions we have to make on the way to simplify the noise term in  the Langevin equation and/or to apply the probabilistic interpretation, therefore, we can consider both of these approaches in the present form only as the QCD motivated models.  }


\preprint{ \bf TAUP - 2854-07\\
\today\\
}
\begin{document}

\def\thefootnote{\arabic{footnote}}
\section{Introduction}
\label{sec:Int}
The simplest approach that we can propose for  high energy interaction is
based \cite{GLR,MUQI} on the BFKL Pomeron \cite{BFKL} and reggeon-like diagram technique
for the BFKL Pomeron interactions \cite{BART,BRN,NP,BLV}. This technique, which is a
generalization of Gribov Reggeon Calculus \cite{GRC},  can be written
in the elegant form of the functional integral (see  \cite{BRN} and the next
section). It is a challenge to solve this theory in QCD  finding the high energy
asymptotic behaviour.  However, even this simple approach has not been solved during
three decades of attempts by the high energy community. This failure stimulates a search
for deeper understanding of physics which is behind the BFKL Pomeron Calculus.
On the other hand, it has been known for three decades that  Gribov Reggeon Calculus has intrinsic
difficulties \cite{KAN} that are related to the overlapping of Pomerons. Indeed, due to this overlapping
we have no hope that the  Gribov Reggeon Calculus could be correct in describing the ultra high energy
asymptotic behaviour of the amplitude. The way out of these difficulties we see in searching for  a new
approach which will  coincide
with the BFKL Pomeron Calculus at high, but not very high,  energies (our correspondence principle)
but it
will be different in the region of ultra high energies. In a  spirit of the  parton approach we
believe that
this effective theory should be based on the interaction of `wee' partons.  We consider, as an important
step in this direction, the observation that has been made at the end of the Reggeon era
\cite{GRPO,LEPO,BOPO},
  that
the Reggeon Calculus
can be reduced to the Markov process \cite{GARD} for the probability of finding a given
number of Pomerons at fixed rapidity $Y$.  Such an interpretation, if it would be
reasonable in QCD,  can be useful,
 since it allows us to use powerful methods of statistical
physics in our search of  the solution.

The logic and scheme of our approach looks as follows.
The first step is the Leading Log (1/x) Approximation (LLA)  of perturbative QCD in which we sum all
contributions of the order of $ \left(\as\,\ln(1/x) \right)^n$. In the LLA we consider such high
energies that
\beq \label{LLA}
\as\,\ln(1/x)\,\,\approx\,\,1\,;\,\,\,\,\,\,\mbox{while}\,\,\,\,\,\,\,\as\,\,\ll\,\,1
\eeq
It is well known that the LLA approach generates the BFKL Pomeron (see  \cite{BFKL} and the next
section)  which leads to the power-like
increase of the scattering amplitude ($ A \,\propto \,\frac{1}{x^{\omega(n=0,\nu=0)}}$ with
 $
\omega(n=0,\nu=0)\,\,\propto\,\as$).

The second step is the BFKL Pomeron Calculus in which we sum all contributions of the order of
\beq \label{PC}
\left( \frac{\bas^2}{N^2_c}\,\frac{1}{x^{\omega(n=0,\nu=0)}} \right)^n
\,\,\,\,\,\,\mbox{therefore}\,\,\,\,\,\,\,
\frac{\bas^2}{N^2_c}\,\frac{1}{x^{\omega(n=0,\nu=0)}}\,\,\approx\,\,1
\eeq
where $\bas\,\,=\,\,\as N_c/\pi$.

The structure of this approach as well as its  parameter has been understood before QCD \cite{VEN}
and was confirmed in QCD (see  \cite{GLR,MUQI,BART,BRN,NP,BLV,IM,KOLE}). This calculus extends the
region of energies from $\ln (1/x) \approx\,1/\bas$ of LLA to $\ln (1/x)
\approx\,(1/\bas)\ln(N^2_c/\bas)$. The BFKL Pomeron Calculus describes correctly the scattering
process in the region of energy:
\beq \label{REPC}
\frac{1}{\bas}\,\,\ln\left(\frac{N^2_c}{\bas}\right)\,\,\ll\,\,ln
\frac{1}{x}\,\,\ll\,\,\,\frac{1}{\bas^2}
\eeq
For higher energies the corrections of the order of $\left(\bas^2 \,\ln (1/x)\right)^n$ should be taken
into account making all calculations very complicated.

Our credo is that we will be able to  describe the high energy processes outside of the region
of \eq{REPC}, if we
could  find an effective theory which  describes the BFKL Pomeron calculus in
the kinematic region given by \eq{REPC}, but based on the microscopic degrees of freedom and  not on the
BFKL Pomeron.  In so doing, we hope that we can avoid all intrinsic difficulties of the  BFKL Pomeron
calculus and build an approximation that will be in an agreement with all general theorems like the
Froissart bounds and so on. Solving this theory,  we can create a basis for moving forward considering
all corrections to this theory due to higher orders in $\bas$ contributions, running QCD coupling and
others.

The goal of this paper is to consider the  key problem: the probabilistic
interpretation of the  BFKL Pomeron Calculus based on the idea that colourless  dipoles
are the correct degrees of freedom in  high energy QCD \cite{MUCD}.
 We believe that colourless dipoles
and their interaction will lead to a future theory at high energies which will have the BFKL Pomeron
Calculus as the low energy limit (see \eq{REPC})  and which will allow us to avoid all difficulties of
dealing with BFKL
Pomerons at ultra high energies.

Colourless dipoles play two different roles in our approach. First, they are
partons (`wee' partons) for the BFKL Pomeron. This role is not related to the large
$N_c$ approximation and, in principle, we can always speak about probability to
find a definite number of dipoles instead of defining the probability to find a
number of the BFKL Pomerons. The second role of the colour dipoles is that at high
energies we can interpret the vertices of Pomeron merging and splitting in terms
of probability for two dipoles  to merge into  one dipole and of probability for
decay of one dipole into two ones. It was shown in \cite{MUCD} that $P
\rightarrow 2P$ splitting can be described as the process of the dipole decay into
two dipoles. However, the relation between the Pomeron merging ($2 P \rightarrow
P$) and the process of merging of two dipoles into one dipole is not so
obvious
and it will be discussed here.

 This paper is a next step in our programme of searching the simplest but
correct approach to high energy scattering in QCD in which we continue the line of thinking presented in
 \cite{L1,L2,L3,L4}.
The outline of the paper looks as follows.

 In the next section we will discuss the
BFKL Pomeron Calculus in the elegant form of the functional integral,  suggested by
M. Braun about five years ago \cite{BRN}.
In the framework of this approach we find a set of equations for the amplitude
of $n$-dipole interaction with the target. We show that the  recent intensive
 work on this subject \cite{IT,MSW,L3} confirms the BFKL Pomeron Calculus in
spite of the fact that these attempts were based on slightly different assumptions.

In section three we demonstrate statistical interpretation of the theory with
 interacting Pomerons. The one-to-one correspondence between BFKL Pomeron calculus and
 Langevin theory is found by showing how full Lagrangian generates stochastic Langevin
 equation with a peculiar noise term. In toy model with zero transverse dimensions this noise
 is reduced to one typical for directed percolation.
 Unfortunately the complexity of the noise term restricts the practical use of Langevin equation
 in this form,  and one should look for further simplifications.
 One of them is to assume that impact parameter is much larger than any dipole size in the system (see \eq{mom3}).
 Using this assumption and going to momentum space we rewrite our theory in the form of Langevin equation with
 a noise term proportional to the field (directed percolation universality class). It should be mentioned
 that large impact parameter approximation is unapplicable for  summation of Pomeron loops, where we integrate over all
 possible values of the impact parameter.

 Next, we discuss an approach based on generating functional. We show the equivalence between
 generating functional approach and
 the BFKL Pomeron calculus in the kinematical region \eq{REPC} that leads to a clear interpretation of
 the BFKL Pomeron calculus as an alternative description of the system of interacting colourless dipoles.
 The interrelation between vertices of the Pomeron interactions and the microscopic dipole processes is considered.
 It is instructive to notice that the
generating functional  approach leads to a feedback to the  BFKL Pomeron Calculus
 restricting the integration over Pomeron fields in the functional integral  by the range $0<\Phi<1/4\pi\as$.

In the  fourth section we suggest a practical way of building  the Monte Carlo code
to solve the equation for the generating functional which can be a basis for
 consideration of the multiparticle production processes.


In conclusion we are going to compare our approach with other approaches on the
market.

\section{The BFKL Pomeron Calculus}
\label{sec:Pomcal}
\subsection{The general structure of the BFKL Pomeron calculus}
We start with a general structure of the BFKL Pomeron calculus in QCD. The BFKL Pomeron exchange can
be written in the form (see \fig{bfklcal}-1)

\beq \label{BFKLC1}
A( \mbox{\fig{bfklcal}-1})\,\,=\,\,\,V_u \bigotimes G_P(r_1,r_2;b|Y_1 - Y_2)\bigotimes
V_d\,\,\propto\,\,\frac{\bas^2}{N^2_c}\,\exp\left(\omega(n=0,\nu=0)\,(Y_1 - Y_2)\right)
\eeq
with $\omega(n=0,\nu=0)\,\propto \,\bas$,$Y_1 - Y_2 = \ln(1/x)$  and $\bigotimes$ denotes the all needed integrations.

It is easy to understand the main parameters of the BFKL Pomeron calculus  by comparing the contributions of the first `fan' diagrams of  \fig{bfklcal}-2 with the one BFKL Pomeron exchange.

This diagram has the following contribution
\bea \label{BFKLC2}
A( \mbox{\fig{bfklcal}-2})&=&\int^{Y_1}_{Y_2}\,d Y' \, V_u\,\bigotimes\, G_P(r_1,r';b | Y_1
-
Y')\,\bigotimes \,\Gamma(1 \to 2)\,
 G^2_P(r',r_2;b | Y' - Y_2)\bigotimes
V^2_d\ \\
&\propto& \frac{V_u\,V^2_d\,\Gamma(1 \to
2)}{\omega(n=0,\nu=0)}\,\exp\left(2\,\omega(n=0,\nu=0)\,
(Y_1 - Y_2)\right)\,\,\propto\,\left(\frac{\bas^2}{N^2_c}\right)^2\,\,\exp\left(2\,\omega(n=0,\nu=0)\,Y_1 - Y_2\right)\nonumber
\eea
where $r_1$ and $r_2$ are the sizes of the projectile and target dipoles while
$\{r'\}$ denotes all
dipole variables in Pomeron splitting and/or merging.

One can see that the ratio of this two diagrams is proportional to $\frac{\bas^2}{N^2_c}\,\exp \left(
\omega(n=0,\nu=0)(Y_1 - Y_2)\right)$ which is the parameter given by \eq{PC}. When this ration is
about 1 we need to calculate all diagrams with the Pomeron exchange and their interactions (see
\fig{bfklcal}-a - \fig{bfklcal}-f ). All vertices, that are shown in \fig{bfklcal},   has been
calculated in  \cite{BART,BRN} and they have the following order in $\bas$ \footnote{In
\eq{VRTS} we use the normalizations of these vertices which are  originated from calculation of the
Feynman diagrams. In  the dipole approach we use a different normalization
(see below section 3 and 4) but all conclusions do not depend on the normalization.}:
$$
\omega(n=0,\nu=o)\,\,\,\propto\,\, \,\bas\,\,; \,\,\,\,\,\,\,\,\Gamma( 1 \to 2)\,\,\,\propto\,\,
\,\frac{\bas^2}{N_c}\,\,; \,\,\,\,\,\,\,\,\,\Gamma( 2 \to
1)\,\,\,\propto\,\,\,\frac{\bas^2}{N_c}\,\,;$$
\beq  \label{VRTS}
\Gamma( 2 \to 2)\,\,\,\propto\,\,\,\frac{\bas}{N^2_c}\,\,;\Gamma( 2 \to
3)\,\,\,\propto\,\,\,\frac{\bas^2}{N^2_c}\,\,; V_u\,\,\propto\,\,\,\frac{\bas}{N_c}\,\,;
V_d\,\,\propto\,\,\,\frac{\bas}{N_c}\,;
\eeq
Using \eq{VRTS} we can easily estimate the contributions of all diagrams in \fig{bfklcal}. Namely,
\bea
A(\mbox{\fig{bfklcal}-3})\,\,\,\propto\,\,\, \,\,L^2\left(Y\right)\,\, ;& &
A(\mbox{\fig{bfklcal}-4})\,\,\,\propto\,\,\, \,\,\frac{\bas}{N^2_c}\,Y\,L^2\left(Y\right)\,\,;\label{diapc4}\\
A(\mbox{\fig{bfklcal}-5})\,\,\,\propto\,\,\, \,\,\frac{1}{N^2_c}\,\,L^3\left(Y\right);& &
A(\mbox{\fig{bfklcal}-6)}\,\,\,\propto\,\,\, \,\,\frac{\bas}{N^2_c}\,Y\,\,L^2\left(Y\right);\label{diapc6}
\eea
with $L\left(Y\right)\,\,=\,\,\left(\bas^2/N^2_c\right)\,\,\exp \left( \omega(n=0,\nu=0)\,Y \right)$.

As we have mentioned in the introduction the BFKL calculus sums all diagrams at such a  high energy that
 parameter $L(Y) $ is of the order of 1 (see \eq{PC}). In this kinematic region we need to take into
account the diagrams of \fig{bfklcal}-1 ,  \fig{bfklcal}-2  and   \fig{bfklcal}-3 (see
\eq{BFKLC1},\eq{BFKLC2} and \eq{diapc4}). Indeed, diagrams of  \fig{bfklcal}-4 and  \fig{bfklcal}-6
( see  \eq{diapc4} and  \eq{diapc6})
are small since $(\bas^2/N^2_c)\,Y \,\,\ll \,\,1/N^2_c\,\,\ll\,\,1$ in the kinematic region of
\eq{REPC}, while the diagrams of  \fig{bfklcal}-5 (see \eq{diapc6}) are small at $L\left(Y\right)
\approx 1$.

The first conclusion that we can derive from this analysis that in the kinematic region where
$L\left(Y\right)\,\,\approx\,\,1$ we need to take into account all diagrams with $\Gamma(1 \to 2) $
and $\Gamma( 2 \to 1)$ vertices while the diagrams with $\Gamma( 2 \to 2)$ and $\Gamma(2 \to 3)$
vertices give small, negligible contributions.

However, if $ L\left(Y\right)\,\,\propto\,\,N_c$ one can see from \eq{BFKLC1} - \eq{diapc6} that all
diagrams give so essential contributions that we have to take them  into account.
Indeed, for such $L(Y)$ $ A(\mbox{\fig{bfklcal}-3})\,\propto\,N^2_c$, $A(\mbox{\fig{bfklcal}-4})\,\propto\, \bas Y
 \,>\, 1$,$ A(\mbox{\fig{bfklcal}-5})\,\propto\,N^2_c$ and $A(\mbox{\fig{bfklcal}-6)}\,\propto \, \bas Y  \,>\, 1$.

It is interesting to  notice that the vertex $\Gamma(2 \to 2) \,\propto\,\,\bas^4/N^2_c$ can be
neglected even at such large values of $ L\left(Y\right)$.

Finally, we can conclude that the first step of our approach can be summing of  the diagrams
with $\Gamma(1 \to 2)$ and $\Gamma(2 \to 1)$ vertices in the kinematic region $L(Y) \approx 1$ or $Y\,\>\,1/\bas\,\ln\Lb N^2_c/\bas\Rb$.

 \FIGURE[ht]{\begin{tabular}{l}
 \centerline{\epsfig{file=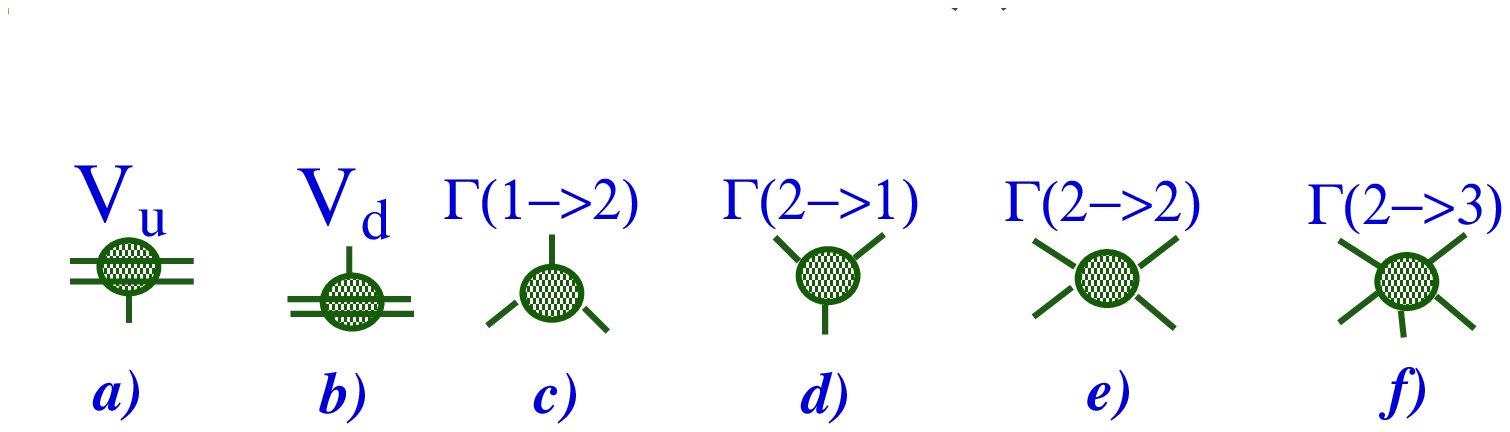,width=160mm}}\\
 \centerline{\epsfig{file=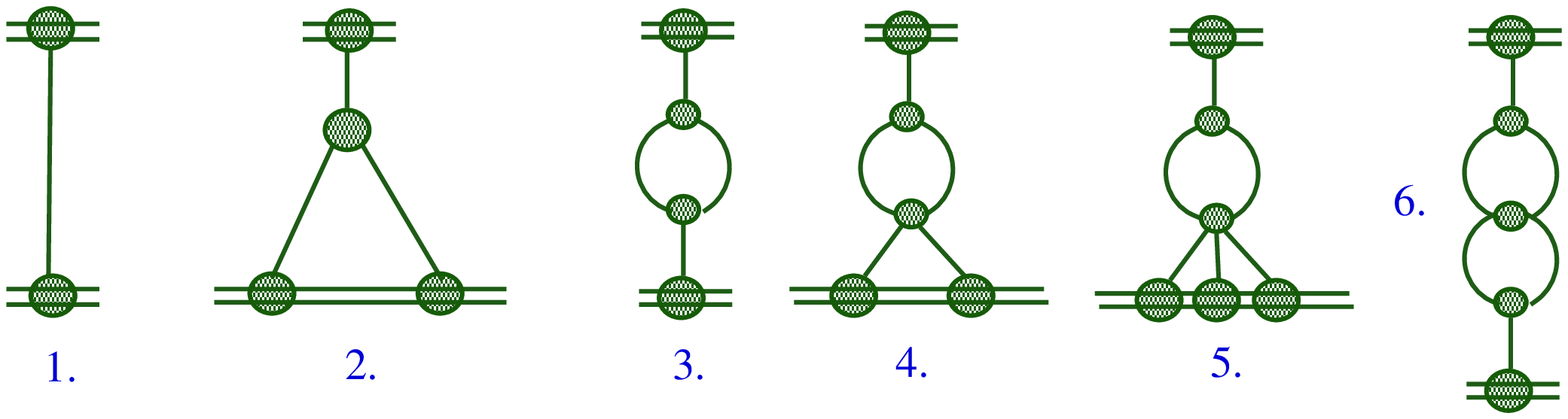,width=160mm}}
 \end{tabular}
 \caption{The BFKL Pomeron interactions and the examples of the
diagrams of the BFKL Pomeron  calculus in QCD. The solid line
describes the Pomeron exchange while the double line stands for the
 dipole.}
 \label{bfklcal}
 }

However, we would like to stress that we need to make an additional assumption inherent for the BFKL
Pomeron calculus: the multi-gluon states in $t$-channel of the scattering amplitude lead to smaller
contribution at high energies than the exchange of the correspondent  number of the BFKL Pomerons (see
more in \cite{L4}). This statement is supported by the fact that numerous attempts to find the intercept of these states being larger than the intercept for multi-Pomeron exchanges\cite{KKM}
have failed.

\subsection{The path integral formulation of the BFKL calculus}
The main ingredient of the  BFKL Pomeron calculus is the Green function  of the BFKL
Pomeron describing the propagation of a pair of gluons from rapidity $Y'$ and
points
$x'_1$ and $x'_2$ to rapidity $Y$ and points $x_1$ and $x_2$ \footnote{Coordinates
$x_i$ here are two dimensional vectors and, strictly speaking, should be denoted by
$\vec{x}_i$ or $\bf{x}_i$. However, we will use notation $x_i$ hoping that it will
not cause difficulties in understanding.}. Since the Pomeron does not carry colour
in $t$-channel we can treat initial and final coordinates as coordinates of quark
and antiquark in a colourless dipole. This Green function is well known\cite{LI},
and has a form
\begin{equation} \label{BFKLGF}
G(x'_1,x'_2;Y | x_1,x_2;Y')\,\,=\,\,\Theta(Y - Y')\times
\end{equation}
$$
\times\,\sum^{\infty}_{n=-\infty} \,\int\,d \nu\,\,d^2 x_0\,e^{\omega(n,\nu) (Y - Y')}\,\lambda(n,\nu)\,
E_n(x'_1,x'_2;x_0|\nu)\,E_n^*(x_1,x_2;x_0|\nu)
$$
where vertices $E_n$ are given by
\begin{equation} \label{BFKLE}
E_n(x_1,x_2;x_0|\nu)\,=\,
\left(\frac{x_{12}}{x_{10}\,x_{20}}
\right)^h\,\left(\frac{x^*_{12}}{x^*_{10}\,x^*_{20}} \right)^{\tilde{h}}
\end{equation}
where $x_{ik} = x_i - x_k $, $x_i = x_{i,x} + i x_{i,y}$ \footnote{$x_{i,x}$ and
$x_{i,y}$  are components of the two dimensional vector $x_i$ on $x$-axis and $y$-
axis} ,$ x^*_i = x_{i,x} + i x_{i,y} $ ; $h = (1 - n)/2 + i\nu$ and $\tilde{h} = 1
- h^*$. The energy levels $\omega(n,\nu)$  are the BFKL eigen values
\begin{equation} \label{BFKLOM}
\omega(n,\nu)\,=\,\bar{\alpha}_S \left( \psi(1) - Re{\, \psi\left(\frac{|n| + 1}{2} + i
\nu\right)}
\right)
\end{equation}
where $\psi(z) = d \ln \Gamma(z)/d z$ and $\Gamma(z)$ is the Euler gamma function.
 Finally
\begin{equation} \label{BFKLLA}
\lambda(n,\nu)\,=\frac{1}{[ ( n + 1)^2 + 4 \nu^2] [(n - 1)^2 + 4 \nu^2]}
\end{equation}


\FIGURE[ht]{
\centerline{\epsfig{file=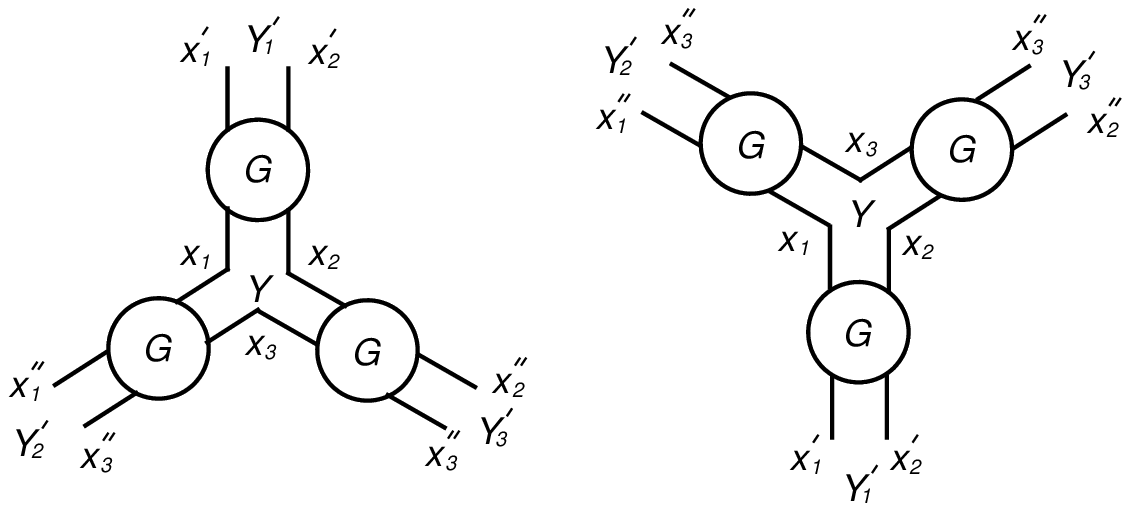,width=120mm}} \caption{The
graphic form of the triple Pomeron vertex in the coordinate
representation.} \label{triple-p} }

The interaction between Pomerons is depicted in Fig. \ref{triple-p}
and  described by the triple Pomeron vertex which can be written in
the coordinate representation \cite{BRN} for the following process:
two gluons with coordinates $x'_1$ and $x'_2$ at rapidity $Y'_1$
decay into two gluon pairs with coordinates $x^{\prime \prime}_1$
and $x^{\prime \prime}_3$ at rapidity $Y'_2$ and $x^{\prime
\prime}_2$ and $x^{\prime \prime}_3$ at rapidity $Y'_3$ due to the
Pomeron splitting at rapidity $Y$. It looks as
\begin{equation} \label{BFKL3P}
2\frac{\pi\,\bar{\alpha}^2_S}{N_c}\,\int\,\frac{d^2 x_1\,d^2\,x_2
\,d^2
x_3}{x^2_{12}\,x^2_{23}\,x^2_{13}}\,\left(G(x'_1,x'_2;Y'_1|x_1,x_2;
Y) \,\,L^{\!\!\!\!\!\!\!\leftarrow}_{1,2} \right)
G(x_1,x_3;Y|x^{\prime\prime}_1,
x^{\prime\prime}_3;Y'_2)\,G(x_3,x_2;Y|x^{\prime\prime}_3,x^{\prime\prime}_2;Y'_3)
 \end{equation}
where
\begin{equation} \label{BFKLL}
L^{\!\!\!\!\!\!\!\leftarrow}_{1,2}\,\,=\,\,r^4_{12}\,p^2_{1}\,p^2_{2}\,\,
\,\mbox{with}\,\,p^2\,=\,-\,\nabla^2
\end{equation}
and arrow shows the direction of action of the operator $L$.
For the inverse process of merging of two Pomerons into one we have
 \begin{equation} \label{BFKLP3}
2\frac{\pi\,\bar{\alpha}^2_S}{N_c}\,\int\,\frac{d^2 x_1\,d^2\,x_2
\,d^2 x_3}{x^2_{12}\,x^2_{23}\,x^2_{13}}\,
G(x^{\prime\prime}_1,x^{\prime\prime}_3;Y'_2|x_1,x_3;Y)\,G(x^{\prime\prime}_3,x^{\prime\prime}_2;Y'_3|x_3,x_2;Y)
\,
\left( L^{\!\!\!\!\!\rightarrow}_{1,2}\, G(x_1,x_2;
Y|x'_1,x'_2;Y'_1) \right)
\end{equation}

The theory with the interaction given by Eq.~(\ref{BFKL3P}) and  Eq.~(\ref{BFKLP3})
can be written through the functional integral  as was proposed and developed by Braun in \cite{BRN}. We include a discussion of the basics of this approach for the sake of completeness of our presentation.
\begin{equation} \label{BFKLFI}
Z[\Phi, \Phi^+]\,\,=\,\,\int \,\,D \Phi\,D\Phi^+\,e^S \,\,\,\mbox{with}\,S \,=\,S_0
\,+\,S_I\,+\,S_E
\end{equation}
where $S_0$ describes free Pomerons, $S_I$ corresponds to their mutual interaction
while $S_E$ relates to the interaction with  external sources (target and
projectile). From Eq.~(\ref{BFKL3P}) and  Eq.~(\ref{BFKLP3}) it is clear that
\begin{equation} \label{S0}
S_0\,=\,\int\,d Y \,d Y'\,d^2 x_1\, d^2 x_2\,d^2 x'_1\, d^2 x'_2\,
\Phi^+(x_1,x_2;Y)\,
G^{-1}(x_1,x_2;Y|x'_1,x'_2;Y')\,\Phi(x'_1,x'_2;Y')
\end{equation}
\beq \label{SI} S_I\,=\,\frac{2\,\pi \bas^2}{N_c}\,\int \,d Y\,\int
\,\frac{d^2 x_1\,d^2 x_2\,d^2 x_3}{x^2_{12}\,x^2_{23}\,x^2_{13}}\,
 \{ \left( L^{\!\!
\!\!\!\rightarrow}_{1,2}\Phi(x_1,x_2;Y)\,\right)\, \,
\Phi^+(x_1,x_3;Y)\,\Phi^+(x_3,x_2;Y)\,\,+\,\,h.c. \} \eeq For $S_E$
we have local interaction both in rapidity and in coordinates,
namely,
 \begin{equation} \label{SE}
S_E\,=\,-\,\int \,dY\,d^2 x_1\,d^2 x_2\, \{
\Phi(x_1,x_2;Y)\,\tau_{pr}(x_1,x_2;Y)\,\,+\,\,\Phi^+(x_1,x_2;Y)\,\tau_{tar}(x_1,x_2;Y)
\}
\end{equation}
where $\tau_{pr}$ ($\tau_{tar}$)  stands for the projectile and target, respectively.
The form of functions $\tau$  depend on the non-perturbative input in our problem
and for the case of nucleus target they are given in \cite{BRN}.

For the case of  projectile being a dipole that scatters off a  nucleus
the scattering amplitude has the form
\begin{equation} \label{T}
T(x_1,y_1;Y)\,\,\equiv
T^{(1)}(x_1,y_1;Y)\,\,=\,\,-\,\frac{4\,\pi^2\,\bar{\alpha}_S}{N_c}\,
\frac{\int\,\,D \Phi\,
D\Phi^+\,\Phi(x_1,y_1;Y)\,e^{S[\Phi,\Phi^+]}}{ \int\,\,D \Phi\, D
\Phi^+\,\,e^{S[\Phi,\Phi^+]}|_{S_E =0}}
\end{equation}
where extra $\alpha_S$ comes from our normalization and  we neglect term with $\tau_{pr}$
in \eq{SE}.

Generally, for the amplitude of interaction of $n$ dipoles at
rapidity $Y$ we can write the following expression
\footnote{Starting from this equation we use notations $x_i$ for the
coordinates of quark while $y_i$ denote the  coordinates of
antiquarks. For  rapidity we will use  $Y$.} \beq \label{TN}
T^{(n)}(x_1,y_1,\dots x_n,y_n;Y)\,\,=\,\,(-1)^n\Lb\,
\frac{4\,\pi^2\,\bar{\alpha}_S}{N_c}\,\Rb^n\, \frac{\int\,\,D \Phi\,
D\Phi^+\,\prod^n_{i=1}\Phi(x_i,y_i;Y)\,\,e^{S[\Phi,\Phi^+]}}{
\int\,\,D \Phi\, D \Phi^+\,\,e^{S[\Phi,\Phi^+]}|_{S_E =0}} \eeq

The extra factor $(-1)^n$ is due to the fact that in $S_E$ the source for both  projectile and
 target, has  extra minus sign.

It is useful to introduce the Green function of the BFKL Pomeron that includes the Pomeron
loops. This function has the form
\beq \label{GF}
G\Lb x_1,y_1;Y|x_2, y_2;Y' \Rb\,\,=\,\,\frac{ \int\,\,D \Phi\,
D\Phi^+\,\Phi^+(x_1,y_1;Y)\,\Phi(x_2,y_2;Y')\,\,e^{S[\Phi,\Phi^+]}}{ \int\,\,D
\Phi\, D \Phi^+\,\,e^{S[\Phi,\Phi^+]}|_{S_E =0}}
\eeq

For further presentation we need some properties of the BFKL Green function
\cite{LI}:

1.  Generally,

\begin{equation}
G^{-1}(x_1,x_2;Y| x'_1,x'_2;Y')\,\, \,=\,p^2_1\,p^2_2\,\left(
\frac{\partial}{\partial Y} + H \right) \,\,=\,\,\left(
\frac{\partial}{\partial Y} + H^+ \right)\,p^2_1\,p^2_2; \label{G1}
\end{equation}

\begin{equation} \label{H}
H f(x_1,x_2;Y) \,\,= \,\,\frac{\bar{\alpha}_S}{2
\pi}\,\int\,\frac{d^2 x_3\,x^2_{12}}{x^2_{23}\,x^2_{13}}\,\left(
f(x_1,x_2;Y)\,-\,f(x_1,x_3;Y)\,-\,f(x_3,x_2;Y) \right) ; \label{G2}
\end{equation}

2. The initial Green function ($G_0$) is
 equal to
\begin{equation} \label{G0}
G_0(x_1,x_2;Y| x'_1,x'_2;Y)\,\,=\,\, \pi^2\,\ln
\frac{x^2_{1,1'}\,x^2_{2,2'}}{x^2_{1,2'}\,x^2_{1',2}}  \,\ln
\frac{x^2_{1,1'}\,x^2_{2,2'}}{x^2_{1,2}\,x^2_{1',2'}}
\end{equation}
 This form of $G_0$ has been discussed in \cite{LI}. In  appendix A we demonstrate that
this expression for $G_0(x_1,x_2;Y| x'_1,x'_2;Y)$ stems from $\omega
= \omega(n=0,\nu)$ term in sum of \eq{BFKLGF}. Only this term is
essential at high energies since all other terms lead to contributions  decreasing
with energy.

3. It should be stressed that
\bea \label{G01}
\nabla^2_1 \,\nabla^2_2 \,G_0(x_1,x_2;Y| x'_1,x'_2;Y)\,\,&=& \\
& & \,\,(2\,\pi)^4\,\,\left(\delta^{(2)}( x_1 - x'_1)\,\delta^{(2)}( x_2
- x'_2)\,+\,\delta^{(2)}( x_1 - x'_2)\,\delta^{(2)}( x_2 -
x'_1)\,\right)\nonumber
\eea

4. In the sum of Eq.~(\ref{BFKLGF}) only the term with $n=0$ is
essential for high energy asymptotic behaviour since all
$\omega(n,\nu) $ with $n   \geq 1$  are negative and, therefore,
lead to contributions that decrease with energy. Taking into account
only the first term one can see that $G$ is the eigen function of
operator $L_{13}$, namely
\beq \label{LG} L_{12}\,G(x_1,x_2;Y|
x'_1,x'_2;Y')\,\,=\,\,\frac{1}{\lambda(0,\nu)} \,G(x_1,x_2;Y|
x'_1,x'_2;Y')\,\,\approx\,\,G(x_1,x_2;Y| x'_1,x'_2;Y') ; \label{L13}
\eeq

 The last equation holds only approximately in the region where
$\nu \,\ll\,1$, but this is the most interesting region which is
responsible for high energy asymptotic behaviour of the scattering
amplitude.

All properties of the BFKL Pomeron Green function as well as of the functional integral approach to the BFKL Pomeron calculus have
 been discussed (for more information see  \cite{BRN,NP,LI}).

  In the next section we will derive  the chain of equations for multi-dipole amplitude in the BFKL Pomeron calculus and will
  show that these equations are the same as ones   that have  been discussed in framework of dipole approach
\cite{IT,MSW,L3}.

\subsection{The chain of equations for the multi-dipole amplitudes}

Using  Eq.~(\ref{BFKLFI}) and Eq.~(\ref{T})
 we can easily obtain the chain equation for multi-dipole amplitude $T^{(n)}$
noticing that every dipole interacts only with one Pomeron (see Eq.~(\ref{T})).

These equations follow from the fact that a change of variables does not alter the value of
functional integral of \eq{BFKLFI}. In particular, $Z[\Phi,\Phi^+] \,=\,Z[\Phi,\Phi'^+]$ (see
\eq{BFKLFI}) where $\Phi'^+ \,=\,\Phi^+ \,+\,\epsilon(x,y)$ with  a  small function $\epsilon(x,y)$.
Therefore,
\beq \label{VZ1}
\int \,D\,\Phi\,D\,\Phi^+ \,e^{S[\Phi\,,\Phi^+]}\,\,=\,\,\int \,D\,\Phi\,D\,\Phi'^+
\,e^{S[\Phi\,,\Phi'^+]}
\eeq
Substituting $\Phi'^+ \,=\,\Phi^+ \,+\,\epsilon(x,y)$ and expanding this equation to first order
in $\epsilon$, we find

\beq \label{VZ2}
0\,=\,\int\,D\,\Phi\,D\,\Phi^+\,e^{S[\Phi\,,\Phi^+]}\,\times \eeq
$$
[\,\int\,d Y \,d Y'\,d^2 x_1\, d^2 x_2\,d^2 x'_1\, d^2 x'_2\,
\epsilon(x_1,x_2;Y)\,
G^{-1}(x_1,x_2;Y|x'_1,x'_2;Y')\,\Phi(x'_1,x'_2;Y')
$$
$$
+\frac{2\,\pi \bas^2}{N_c}\,\int \,d Y\,\int \,\frac{d^2 x_1\,d^2
x_2\,d^2 x_3}{x^2_{12}\,x^2_{23}\,x^2_{13}}\, \cdot \{
 \left( \epsilon(x_1,x_2;Y)L^{\!\! \!\!\!\leftarrow}_{1,2}\,\right)\,\cdot\,
\Phi(x_1,x_3;Y)\,\Phi(x_3,x_2;Y)   +$$
$$
  2\left( L^{\!\!
\!\!\!\rightarrow}_{1,2}\Phi(x_1,x_2;Y)\,\right)\,\cdot\,
\epsilon(x_1,x_3;Y) \,\Phi^+(x_3,x_2;Y)
  \} -\,\int \,dY\,d^2 x_1\,d^2 x_2\,
  \epsilon(x_1,x_2;Y)\,\tau_{tar}(x_1,x_2;Y)]
$$

We redefine the integration variables in the third term as follows
$$
2\left( L^{\!\!
\!\!\!\rightarrow}_{1,2}\Phi(x_1,x_2;Y)\,\right)\,\cdot\,
\epsilon(x_1,x_3;Y) \,\Phi^+(x_3,x_2;Y) \longrightarrow 2\left(
L^{\!\! \!\!\!\rightarrow}_{1,3}\Phi(x_1,x_3;Y)\,\right)\,\cdot\,
\epsilon(x_1,x_2;Y) \,\Phi^+(x_2,x_3;Y)
$$

Using the expression for the Hamiltonian Eq. \ref{G1} and the
Casimir operator Eq. \ref{BFKLL} we define a new variation parameter
$\epsilon(x_1,x_2;Y)p^2_{1}\,p^2_{2}$. In terms of this variation
parameter Eq. \ref{VZ2} reads as
 \beq \label{VZ2-a1}
0\,=\,\int\,D\,\Phi\,D\,\Phi^+\,e^{S[\Phi\,,\Phi^+]}\,\times \eeq
$$
[\,\int\,d Y  \,d^2 x_1\, d^2 x_2 \;\; \epsilon(x_1,x_2;Y)p^2_1p^2_2
\Lb\frac{\partial}{\partial Y}+ H \Rb\Phi(x_1,x_2;Y)
$$
$$
+\frac{2\,\pi \bas^2}{N_c}\,\int \,d Y\,\int \,\frac{d^2 x_1\,d^2
x_2\,d^2 x_3}{x^2_{12}\,x^2_{23}\,x^2_{13}}\, \cdot \{
 \left( \epsilon(x_1,x_2;Y)p^2_1 p^2_2\right)\,\cdot\,
\Phi(x_1,x_3;Y)\,\Phi(x_3,x_2;Y)   +$$
$$
  2\left( L^{\!\!
\!\!\!\rightarrow}_{1,3}\Phi(x_1,x_3;Y)\,\right)\,\cdot\,
\epsilon(x_1,x_2;Y) \, \frac{p^2_1p^2_2}{p^2_1p^2_2}
\,\Phi^+(x_3,x_2;Y)
  \} -\,\int \,dY\,d^2 x_1\,d^2 x_2\,
  \epsilon(x_1,x_2;Y)\frac{p^2_1p^2_2}{p^2_1p^2_2}\,\tau_{tar}(x_1,x_2;Y)]
$$

We denote the new variation parameter by
$\tilde{\epsilon}(x_1,x_2;Y) =p^2_1p^2_2\epsilon(x_1,x_2;Y)$ and use
the property of the initial Green function Eq. \ref{G01} to rewrite
$\frac{1}{p^2_1p^2_2}$ in terms of $G_0$ as follows
 \beq \label{I0}
 \frac{1}{p^2_1p^2_2}\tilde{\epsilon}(x_1,x_2;Y)\,\,=\,\,
 \frac{1}{2\,(2\pi)^4}\int G_0(x_1,x_2;Y| x'_1,x'_2;Y)\tilde{\epsilon}(x'_1,x'_2;Y)
 d^2x'_1 d^2x'_2
\eeq
Thus, Eq. \ref{VZ2-a1} can be written as
 \beq \label{VZ2-a2}
0\,=\,\int\,D\,\Phi\,D\,\Phi^+\,e^{S[\Phi\,,\Phi^+]}\,\times \eeq
$$
[\,\int\,d Y  \,d^2 x_1\, d^2 x_2 \;\; \tilde{\epsilon}(x_1,x_2;Y)
\Lb\frac{\partial}{\partial Y}+ H \Rb\Phi(x_1,x_2;Y)
$$
$$
+\frac{2\,\pi \bas^2}{N_c}\,\int \,d Y\,\int \,\frac{d^2 x_1\,d^2
x_2\,d^2 x_3}{x^2_{12}\,x^2_{23}\,x^2_{13}}\, \cdot \{
  \tilde{\epsilon}(x_1,x_2;Y) \, \,
\Phi(x_1,x_3;Y)\,\Phi(x_3,x_2;Y)   +$$
$$
  2\left( L^{\!\!
\!\!\!\rightarrow}_{1,3}\Phi(x_1,x_3;Y)\,\right)\,\cdot\,
\left\{\frac{1}{2\,(2\pi)^4}\int G_0(x_1,x_2;Y|
x'_1,x'_2;Y)\tilde{\epsilon}(x'_1,x'_2;Y)
 d^2x'_1 d^2x'_2 \right\}
\,\Phi^+(x_3,x_2;Y)
  \}
$$
$$
   -\,\int \,dY\,d^2 x_1\,d^2 x_2\,
  \left\{\frac{1}{2\,(2\pi)^4}\int G_0(x_1,x_2;Y|
x'_1,x'_2;Y)\tilde{\epsilon}(x'_1,x'_2;Y)
 d^2x'_1 d^2x'_2 \right\}\,\tau_{tar}(x_1,x_2;Y)]
$$

 Noting that the r.h.s. of \eq{VZ2-a2} should vanish for any
possible variation of  $\tilde{\epsilon}(x_1,x_2;Y)$
 we obtain

 \beq \label{VZ2-a3}
0\,=\,\int\,D\,\Phi\,D\,\Phi^+\,e^{S[\Phi\,,\Phi^+]}\,\times [
\Lb\frac{\partial}{\partial Y}+ H \Rb\Phi(x_1,x_2;Y)  \eeq
 $$
+\frac{2\,\pi \bas^2}{N_c}\, \,\int \,\frac{x^2_{12}\,d^2
x_3}{x^2_{23}\,x^2_{13}}
 \Phi(x_1,x_3;Y)\,\Phi(x_3,x_2;Y)   +$$
$$
+\frac{2\,\pi \bas^2}{N_c}\,\frac{ 2}{(2\pi)^4} \,\int \,
\frac{\,d^2x'_1\,d^2x'_2\,d^2x_3}{x^2_{1'2'}x^2_{2'3}\,x^2_{1'3}}
 \left( x^4_{1' 3}\;p^2_{1'}p^2_3\Phi(x_1',x_3;Y)\,\right)
   G_0(x'_1,x'_2;Y|
x_1,x_2;Y) \,\Phi^+(x_3,x'_2;Y)
  \}
$$
$$
   -\frac{1}{2\,(2\pi)^4} \,\int d^2x'_1 d^2x'_2
    G_0(x'_1,x'_2;Y|
x_1,x_2;Y)
  \,\tau_{tar}(x'_1,x'_2;Y)]
$$
where  we interchanged
$(x'_1,x'_2)\leftrightarrow (x_1,x_2)$. We notice that the third and
last terms are independent of rapidity and  can be absorbed  in  the initial
condition. This is obvious for the last term which represents the
target source. To show this for the third term we use the property
of the Casimir operator at high energies ($n=0\;,\;\nu=0$)
$$
 L^{\!\!
\!\!\!\rightarrow}_{1,3}\Phi(x_1,x_3;Y)\simeq \Phi(x_1,x_3;Y)
$$
and the definition of the Green function (see Eq. \ref{GF}). This equation will be discussed in the next section in more details.
 We see
that the third term results into the product of two initial Green
functions which are independent of rapidity.

Now we can use the definition of the amplitude defined in Eq.
\ref{T} and Eq. \ref{TN} to rewrite Eq. \ref{VZ2-a3} in a simple
form
\begin{equation} \label{T1-chain}
\frac{\partial T^{(1)}(x_1,x_2;Y)}{\partial Y}\,\,=\,\,
\frac{\bar{\alpha}_S}{2 \,\pi}\,\int\,d^2\,z\,K(x_1,x_2;z)\,
\end{equation}
$$
\left( T^{(1)}(x_1,z;Y)\,+\,T^{(1)}(z,x_2;Y)\, -\,
T^{(1)}(x_1,x_2;Y)\,-\,T^{(2)}(Y;x_1,z;z,x_2;Y)\right)
$$
where kernel $K(x,y|z)$ is defined as
\begin{equation} \label{K}
K(x,y|z)\,\,=\,\,\frac{(x - y)^2}{(x - z)^2\,(z - y)^2}
\end{equation}
and the Hamiltonian is given by Eq. \ref{G2}.

This equation has a very simple meaning that is clear from
\fig{pomeq}.
 \FIGURE[ht]{
\centerline{\epsfig{file=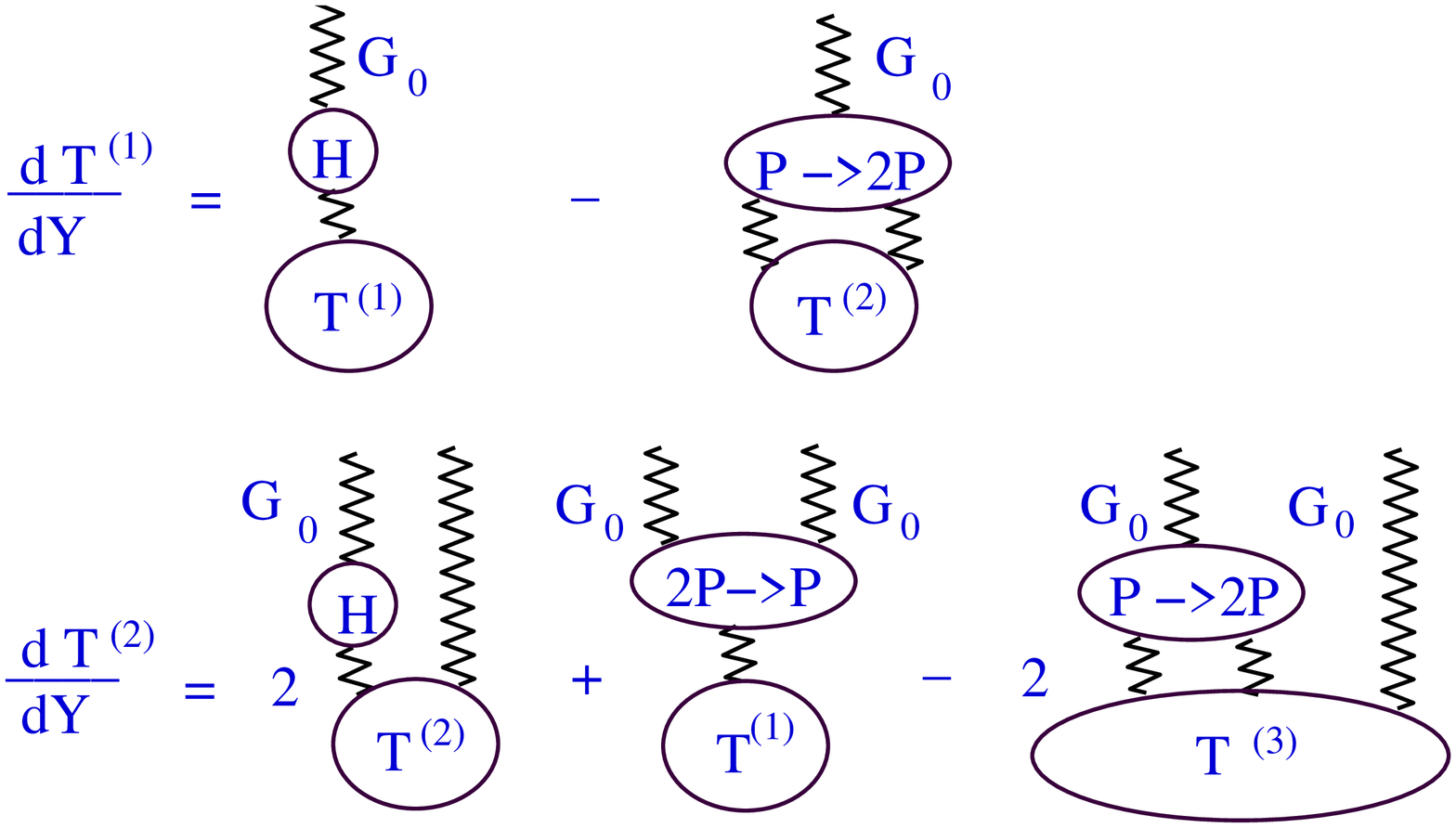,width=120mm}} \caption{The
graphic form of equations for the multi dipole amplitude.}
\label{pomeq} }

Starting from equation \beq \label{VZ4} \int \,D\,\Phi\,D\,\Phi^+
\,\Phi(Y; x_4,x_5)\,\, e^{S[\Phi\,,\Phi^+]}\,\,=\,\,\int
\,D\,\Phi\,D\,\Phi'^+\,\Phi(Y; x_4,x_5)\, \,e^{S[\Phi\,,\Phi'^+]}
\eeq we obtain the equation for the amplitude $T^{(2)}$, namely,
\begin{equation}  \label{T2-chain}
\frac{\partial T^{(2)}(x_1,x_2;x_3,x_4;Y)}{\partial Y} \,\,=\,\,
\frac{\bar{\alpha}_S}{2 \,\pi}\int\,d^2\,z\,K(x_1,x_2|z)\,
\end{equation}
$$
\left(  T^{(2)}(x_1,z;x_3,x_4;Y) \, +\,T^{(2)}(z,x_2;x_3,x_4;Y)
\,-\, \,T^{(2)}(x_1,x_2;x_3,x_4;Y)\, - \right.
$$
$$
\left. -\,T^{(3)}(x_1,z;z,x_2;x_3,x_4;Y) \right)\,+
$$
$$
+\,   \frac{\bas^2}{N^2_c} \, \frac{\bar{\alpha}_S}{2 \,\pi} \int
\,d^2\,x'\,d^2\,x^{\prime \prime}\,\Gamma_{2 \to 1}(x_1,x_2;x_3,x_4|
x',x^{\prime \prime})\,\nabla^2_{x'}\nabla^2_{x^{\prime \prime}}
T^{(1)}(x',x^{\prime \prime};Y);
$$
 where function $\Gamma_{2\to 1}$ is equal to
\begin{equation} \label{V21}
\Gamma_{2 \to 1}(x_1,y_1;x_2,y_2|x,y)\,\,= \int\,d^2 z
K(x,y;z)\,\,G_0(x_1,y_1|x,z)\,G_0(x_1,y_1|z,y)
\end{equation}
In derivation of  Eq.~(\ref{T1-chain}) and Eq.~(\ref{T2-chain}) we use
Eq.~(\ref{G0}) and Eq.~(\ref{G01}) as well as normalization
condition (see Eq.~(\ref{T})) for the scattering amplitude. These
two equations are the same as in \cite{IST}.
This shows that approach developed in \cite{IT,IST} gives a result similar to that obtained
from BFKL Pomeron Calculus.

Assuming  $T^{(2)}= T^{(1)}\,T^{(1)}$ we obtain the Balitsky-Kovchegov equation
\cite{B,K}. We can do this only if we can argue why the Pomeron splitting is more
important than the Pomeron merging. For example,  this assumption is reasonable for
scattering of the dipole with the nucleus target.   Generally speaking, the splitting
and merging have the same order in $\alpha_S$ ( see Eq.~(\ref{BFKL3P}) and
 Eq.~(\ref{BFKLP3}) ). In  Eq.~(\ref{T1-chain}) and  Eq.~(\ref{T2-chain}) these two processes  look
like having a  different order of magnitude in $\alpha_S$, but this fact does not
interrelate with any physics and reflects  only our normalization. However, we will
see that for a  probabilistic interpretation the correct normalization is very
important.

\section{Generating functional and probabilistic interpretation}
\label{sec:GenFPI}
\subsection{Statistical physics analogy: Langevin equation and directed percolation}

The functional of \eq{BFKLFI} is reminiscent of the partition function of statistical mechanics.
Indeed, the partition function has a general form
\beq \label{SM1}
Z[H]\,\,=\,\,e^{ - \frac{1}{kT}\,F[H]}\,\,=\,\,\int d {\bf s}\,e^{ - \frac{1}{kT}\,\int\,d x \Lb {\cal
H}(s)\,- H \,s(x)\Rb}
\eeq
where $F[H]$ is the Helmholtz free energy.  As an example, \eq{SM1} is written for the system of spins
with the energy density ${\cal H}(s)$
in the external magnetic field $H$. The integration is performed over all possible spin configuration in the system.

Comparing \eq{SM1} and \eq{BFKLFI} one can see that \eq{BFKLFI} describes a statistical system with
$kT=1$ and  with $F[-\tau_{tr}] = - S$. The form of $S_E$ suggests that $H = - \tau_{tr}$ plays a role
of the external field.

We demonstrate the idea of the statistical interpretation of our theory by showing how one can
generate full interaction Lagrangian starting with Langevin equation with a specific choice of a noise term.

We define a partition function as an integral over all possible
field configurations being restricted to satisfy Langevin equation with a Gaussian noise
\begin{eqnarray} \label{noise1}
&&Z[\Phi,\zeta] \sim \int P(\zeta) \; D \zeta \;  D \Phi \times  \\
&&\delta  \left(\int d^2x_1 d^2x_2 d Y \{\frac{\partial \Phi (x_1,x_2;Y) }{\partial Y} + H \Phi (x_1,x_2;Y)+
\frac{2\pi \bar{\alpha}_S}{N_c}\int \frac{x^2_{12}d^2 x_3}{x^2_{23}x^2_{13}}\Phi(x_1,x_3;Y)\Phi(x_3,x_2;Y)
+\zeta(x_1,x_2;Y) \}
 \right)\nonumber
\end{eqnarray}
where $P(\zeta)$ stands for Gaussian probability distribution of the noise with $<\zeta(x_1,x_2;Y)>=0$.
The noise $\zeta(x_1,x_2;Y)$ has correlation function

\begin{eqnarray}\label{corr1}
<\zeta(x_1,x_2;Y)\zeta(x^{\prime}_1,x_2^{\prime};Y^{\prime})>=B\; \delta(Y-Y^{\prime}) \;
 \delta(x_1-x_1^{\prime})\;
 \delta(x_2-x_2^{\prime})
 \end{eqnarray}

where $B$ is some function of $x_1$, $x_2$ and $Y$, and will be determined later.

We want to rewrite \eq{noise1} as a functional integral with a term responsible for merging of Pomerons.
To do this we introduce a response field $\tilde{\Phi}(x_1,x_2;Y)$ and use exponential representation
of $\delta$-function
\begin{eqnarray} \label{noise2}
Z[\Phi,\tilde{\Phi},\zeta] &&\sim \int P(\zeta) \; D \zeta \;  D \Phi \; D \tilde{\Phi}\;\;
 \exp ( i \int d^2x_1 d^2x_2 d Y \;  \tilde{\Phi}(x_1,x_2;Y)\{\frac{\partial \Phi (x_1,x_2;Y) }{\partial Y}
 + H \Phi (x_1,x_2;Y) \nonumber \\
 &&+ \frac{2\pi \bar{\alpha}_S}{N_c}\int \frac{x^2_{12}d^2 x_3}{x^2_{23}x^2_{13}}\Phi(x_1,x_3;Y)\Phi(x_3,x_2;Y)
+\zeta(x_1,x_2;Y) \})
\end{eqnarray}
Performing functional integration over noise $\zeta(x_1,x_2;Y)$ first, we have
\begin{eqnarray} \label{noise22}
Z[\Phi,\tilde{\Phi}] &&\sim \int \;  D \Phi \; D \tilde{\Phi}\;\;
 \exp ( i \int d^2x_1 d^2x_2 d Y \;  \tilde{\Phi}(x_1,x_2;Y)\{\frac{\partial \Phi (x_1,x_2;Y) }{\partial Y}
 + H \Phi (x_1,x_2;Y) \nonumber \\
 &&+ \frac{2\pi \bar{\alpha}_S}{N_c}\int \frac{x^2_{12}d^2 x_3}{x^2_{23}x^2_{13}}\Phi(x_1,x_3;Y)\Phi(x_3,x_2;Y)
+\frac{i}{2}B\;\tilde{\Phi}(x_1,x_2;Y)
 \})
\end{eqnarray}
Here we used properties of Gaussian integral.
Now we compare functional integral of \eq{noise2} with one given by \eq{BFKLFI}. It is clear that in order to have
proper free and interaction terms we should identify arbitrary $\tilde{\Phi}(x_1,x_2;Y)$ and $B$ as
\begin{eqnarray}\label{noise3}
i\; \tilde{\Phi}(x_1,x_2;Y)\equiv p^2_1\; p^2_2 \;\Phi^{+}(x_1,x_2;Y)
\end{eqnarray}
and
\begin{eqnarray}\label{noise4}
B\equiv  2\frac{2\pi \bar{\alpha}_S}{N_c}
\left(\frac{1}{p^2_1p^2_2 \Phi^{+}(x_1,x_2;Y)}\right)^2
\int \frac{d^2x_3}{x^2_{12}x^2_{13}x^2_{23}}
(L_{12}\Phi(x_1,x_2;Y)) \Phi^{+}(x_1,x_3;Y)\Phi^{+}(x_3,x_2;Y)
 \end{eqnarray}
 The last missing thing is to consider functional integration
measure. After the redefinition of the response field the measure of
functional integration is $D(-i\;p^2_1p^2_2 \;\Phi^{+}(x_1,x_2;Y))$ instead of desirable $D(\Phi^{+}(x_1,x_2;Y))$.
But according to the definition of the functional integration  we have to sum over all possible field configurations.
After performing change  $D(-i\;p^2_1p^2_2 \;\Phi^{+}(x_1,x_2;Y)) \rightarrow  D(\Phi^{+}(x_1,x_2;Y))$ we still account for all possible configurations up to, maybe, an infinite constant, which cancels in any calculations of  physical
observables.

Thus, using a Gaussian noise of the form \eq{noise4} we fully reproduce functional \eq{BFKLFI}, except the
terms of Lagrangian corresponding to external sources. As it was already mentioned these terms can be absorbed
in the initial condition of Langevin equation.
It should be stressed that correspondence between functional and stochastic formulation of the problem was shown on
general basis without any additional assumptions.

In the toy model of zero transverse dimensions the action is given
by
\begin{eqnarray}\label{toy_S}
S \;=\; \int d Y  \left(\Phi^+(Y)\{\frac{d}{d Y}-\Delta\}\Phi(Y)
+G_{3P}\left(\Phi^+(Y)\Phi^2(Y)+\Phi(Y)\Phi^{+2}(Y)\right) -\Phi^+(Y)\tau_{tr}\delta(Y-Y_0)\right)
\end{eqnarray}
where $\Delta$ and  $G_{3P}$ are Pomeron intercept and triple Pomeron vertex.

 For this form of action \eq{noise4} becomes
\begin{eqnarray} \label{Bzero}
B=2G_{3P}\Phi
\end{eqnarray}
leading to correlation function \eq{corr1} to be proportional to the field. This means that our theory belongs to
direct percolation universality class.

As a final step of our discussion we would like to write Langevin equation used in \eq{noise1}
 in a slightly different way. Because of Gaussian form of probability distribution $P(\zeta)$ we may
 make the following change of variables
 \begin{eqnarray} \label{noise5}
 \zeta(x_1,x_2;Y)\rightarrow \sqrt{B}\;\zeta(x_1,x_2;Y)
 \end{eqnarray}
 In this case corresponding Langevin equation reads
 \begin{eqnarray} \label{noise6}
 \frac{\partial \Phi (x_1,x_2;Y) }{\partial Y} + H \Phi (x_1,x_2;Y)-
\frac{2\pi \bar{\alpha}_S}{N_c}\int \frac{x^2_{12}d^2 x_3}{x^2_{23}x^2_{13}}\Phi(x_1,x_3;Y)\Phi(x_3,x_2;Y)
+\sqrt{B}\zeta(x_1,x_2;Y)=0 \nonumber \\
 \end{eqnarray}
 and correlation function \eq{corr1} is given by
 \begin{eqnarray}\label{corr2}
<\zeta(x_1,x_2;Y)\zeta(x^{\prime}_1,x_2^{\prime};Y^{\prime})>= \delta(Y-Y^{\prime}) \;
 \delta(x_1-x_1^{\prime})\;
 \delta(x_2-x_2^{\prime})
 \end{eqnarray}
 In the toy model the Langevin equation for redefined noise \eq{noise5} is given by
 \begin{eqnarray}\label{noise7}
\frac{\partial \Phi(Y)}{\partial Y}=\Delta \Phi(Y) -G_{3P}\Phi^2(Y) -\sqrt{2G_{3P}\Phi(Y)}\zeta(Y)
 \end{eqnarray}
Therefore, the description of the BFKL Pomeron calculus with the help of the Langevin equation for
directed percolation is equivalent to the functional integral with the action of \eq{BFKLFI}.
In QCD we expect (see \cite{L3,L4}) a different form of correlation function
 ($\;\;<\zeta \zeta> \propto \Phi(1-\Phi) \;\;$) which belongs to universality class different from that of \eq{Bzero}.

Going back to coordinate dependence we must admit that the form of noise \eq{noise4} is too much complicated for
 calculations, and we will use $L_{1,2}\Phi(1,2)\simeq \Phi(1,2)$ approximation for our further considerations.
We believe that this approximation describes well Pomeron dynamics at high energies. The justification of this assumption
is given in the next section, where we consider Langevin equation for a mixed momentum and impact parameter representation.

\subsection{Langevin equation in mixed representation}
 At high energies both $n$ and $\nu$ are small and one can assume that
 \begin{eqnarray}\label{mom1}
 L_{1,2}\Phi(1,2)\simeq \Phi(1,2) \;\;\;  \rightarrow  \;\;\;\;
  \nabla^2_1 \nabla^2_2 \Phi(1,2)\simeq \frac{1}{x^4_{12}}\Phi(1,2)
 \end{eqnarray}
 This assumption means that for all BFKL Pomerons in the BFKL Pomeron calculus we can use the diffusion approximation,
  which is equivalent of using  Fisher-Kolmogorov-Petrov-Piskounov
   equation in statistical approach. In other words, we assume that all
    partial energies in the BFKL Pomeron calculus are high.
  In this case the form of the action of \eq{BFKLFI} simplifies considerably
  \begin{eqnarray}\label{mom2}
  S_0+S_I&=&\int \frac{d^2 x_1 d^2x_2}{x^4_{12}}\Phi^+(x_1,x_2;Y)\frac{\partial \Phi(x_1,x_2;Y)}{\partial Y} \nonumber
  \\
  &&+ \int d Y \frac{d^2x_1 d^2 x_2 d^2 x_3}{x^2_{12} x^2_{23}x^2_{13}} \left( \frac{\bar{\alpha}^{}_S}{2\pi}
 \Phi^+(x_1,x_2;Y)\{\Phi(x_1,x_2;Y)-\Phi(x_1,x_3;Y) - \Phi(x_3,x_2;Y)\} \right.  \nonumber\\
&& \hspace{4cm}\left.
 +\frac{2\pi \bar{\alpha}^2_S}{N_c}\{
 \Phi (x_1,x_2;Y)\Phi^+(x_1,x_3;Y)\Phi^+(x_3,x_2;Y) +h.c.
 \}
 \right)
  \end{eqnarray}

  To justify the  assumption of \eq{mom1} we want to show how the simplified action of \eq{mom2} generates the
  same BFKL calculus as the full one. To see this we consider the enhanced diagram of \fig{bfklcal}-3 using the
  full action  of \eq{BFKLFI}, namely,
  \begin{eqnarray}\label{enh1}
  A(\fig{bfklcal}-3)=\int \frac{d^2x^{\prime}_1 d^2x^{\prime}_2 d^2x^{\prime}_3}
  {x^{\prime 2 }_{12}x^{\prime 2 }_{13}x^{\prime 2 }_{23}}
  \frac{d^2x^{\prime \prime}_1 d^2x^{\prime \prime}_2 d^2x^{\prime \prime}_3}
  {x^{\prime \prime 2 }_{12}x^{\prime \prime 2 }_{13}x^{\prime \prime 2 }_{23}}
  \int_0^Y dY^{\prime} \int_0^{Y^{\prime}} dY^{\prime\prime}
  G(x^p_1,x^p_2;Y|x^{\prime}_1,x_2^{\prime};Y^{\prime}) L^{\!\!\!\!\!\leftarrow}_{1^{\prime},2^{\prime}} \nonumber \\
  \times G(x^{\prime}_1,x_3^{\prime};Y^{\prime}|x^{\prime \prime}_1,x_3^{\prime\prime};Y^{\prime \prime})
  G(x^{\prime}_3,x_2^{\prime};Y^{\prime}|x^{\prime \prime}_3,x_2^{\prime\prime};Y^{\prime \prime})
   L^{\!\!\!\!\!\rightarrow}_{1^{\prime \prime},2^{\prime \prime}}
   G(x^{\prime\prime}_1,x_2^{\prime\prime};Y^{\prime\prime}|x^{t}_1,x_2^{t};0)
   \end{eqnarray}
  \eq{BFKLGF} can be written in the following form
   \begin{eqnarray}\label{enh2}
   G(x^{p}_1,x_2^{p};Y|x^{\prime}_1,x_2^{\prime};Y^{\prime})
   =\sum^{\infty}_{n=-\infty} \int d \nu \int d^2x_0
   e^{\omega(n,\nu)(Y-Y')}G_{n}(x^p_1,x_2^p;x^{\prime}_1,x^{\prime}_2|\nu)
   \end{eqnarray}
   Therefore
   \begin{eqnarray}\label{enh3}
   G(x^{p}_1,x_2^{p};Y|x^{\prime}_1,x_2^{\prime};Y^{\prime})L^{\!\!\!\!\!\leftarrow}_{1^{\prime},2^{ \prime}}
  &=&\sum^{\infty}_{n=-\infty} \int d \nu \int d^2x_0
   e^{\omega(n,\nu)(Y-Y')}G_{n}(x^p_1,x_2^p;x^{\prime}_1,x^{\prime}_2|\nu)
   L^{\!\!\!\!\!\leftarrow}_{1^{\prime},2^{ \prime}}
   \nonumber \\
   &=&\sum^{\infty}_{n=-\infty} \int d \nu \int d^2x_0
   e^{\omega(n,\nu)(Y-Y')}\frac{1}{\lambda(n,\nu)} \;G_{n}(x^p_1,x_2^p;x^{\prime}_1,x^{\prime}_2|\nu)
   \nonumber \\
   &\stackrel{\rm  n=0,\nu \rightarrow 0}{\longrightarrow} &
    \int d \nu \int d^2x_0
   e^{\omega(0,\nu)(Y-Y')}\frac{1}{\lambda(0,0)} \;G_{n=0}(x^p_1,x_2^p;x^{\prime}_1,x^{\prime}_2|\nu)
   \nonumber \\
   &\simeq & G(x^{p}_1,x_2^{p};Y|x^{\prime}_1,x_2^{\prime};Y^{\prime})
      \end{eqnarray}
      where we used the fact that $Y-Y^{\prime}$, $Y^{\prime}-Y^{\prime \prime}$
and $Y^{\prime \prime}-0$ are so large that we can account only for $n=0$ term and replace $\lambda(0,\nu)$ by $1$
(see \eq{BFKLLA}). In Appendix \ref{sec:A} we show that replacing the sum over $n$ in  \eq{enh2} by the $n=0$ term
we obtain the Green function of the BFKL Pomeron that satisfies initial condition determined by the Born approximation.
Using \eq{enh3} we can rewrite \eq{enh1} as
\begin{eqnarray}\label{enh11}
  A(\fig{bfklcal}-3)=\int \frac{d^2x^{\prime}_1 d^2x^{\prime}_2 d^2x^{\prime}_3}
  {x^{\prime 2 }_{12}x^{\prime 2 }_{13}x^{\prime 2 }_{23}}
  \frac{d^2x^{\prime \prime}_1 d^2x^{\prime \prime}_2 d^2x^{\prime \prime}_3}
  {x^{\prime \prime 2 }_{12}x^{\prime \prime 2 }_{13}x^{\prime \prime 2 }_{23}}
  \int_0^Y dY^{\prime} \int_0^{Y^{\prime}} dY^{\prime\prime}
  G(x^p_1,x^p_2;Y|x^{\prime}_1,x_2^{\prime};Y^{\prime})  \nonumber \\
  \times G(x^{\prime}_1,x_3^{\prime};Y^{\prime}|x^{\prime \prime}_1,x_3^{\prime\prime};Y^{\prime \prime})
  G(x^{\prime}_3,x_2^{\prime};Y^{\prime}|x^{\prime \prime}_3,x_2^{\prime\prime};Y^{\prime \prime})
     G(x^{\prime\prime}_1,x_2^{\prime\prime};Y^{\prime\prime}|x^{t}_1,x_2^{t};0)
   \end{eqnarray}
   which is the expression for the diagram of \fig{bfklcal}-3 for the action in \eq{mom2}.

  \eq{mom2} can be reduced to a more elegant form  in mixed momentum and impact parameter representation.
  We assume that impact parameter is much larger than a dipole size
  \begin{eqnarray} \label{mom3}
  \frac{x_1+x_2}{2}\gg x_1-x_2
  \end{eqnarray}
  Indeed, we expect that the typical size of the dipoles will be of the order of  inverse saturation momentum $1/ Q_s(x)$, while the typical impact parameter of the scattering dipole should be much larger (at least of the order of the size of the hadron). Thus, we take $b=(x_1+x_2)/2\simeq (x_1+x_3)/2 \simeq (x_2+x_3)/2$.

  We introduce the Fourier transform
  \begin{eqnarray}\label{mom4}
  \Phi(x_1,x_2;Y)=x^2_{12}\int d^2 k \;e^{i  \vec{k} \cdot \vec{x}_{12}}\; \Phi(k,b;Y)
  \end{eqnarray}
  In terms of fields $\Phi(k,b;Y)$ and $\Phi^+(k,b;Y)$ the action looks as follows

  \begin{eqnarray}\label{mom5}
  S=S_0+S_I=\int d^2k\; d^2b \;dY \; \Phi^+(k,b;Y)&& \left(\frac{\partial \Phi(k,b;Y)}{\partial Y}
  -\frac{\bar{\alpha}^{}_S}{2\pi}\int d^2k^{\prime} K(k,k^{\prime})\Phi(k^{\prime},b;Y)
 \right. \nonumber \\
  &&\left.+
  \frac{2\pi \bar{\alpha}^2_S}{N_c}\{\Phi^+(k,b;Y)\Phi(k,b;Y)+\Phi(k,b;Y)\Phi(k,b;Y)\}
    \right)
 \end{eqnarray}

 where $K(k,k^{\prime})$ is the BFKL kernel in the momentum representation, namely,
 \begin{eqnarray}\label{mom6}
 \int \; d^2 \;   k^{\prime} \; K(k, k^{\prime}) \; \Phi(k^{\prime},b;Y)\; =
 \int d^2k^{\prime} \frac{\Phi(k^{\prime},b;Y)}{(k-k^{\prime})^2}-
 k^2 \int d^2 k  \; k^{\prime} \frac{\Phi(k,b;Y)}{(k-k^{\prime})^2(k^{\prime 2}+(k-k^{\prime})^2)}
 \end{eqnarray}
 In \eq{mom5} we obtained interaction terms multiplied
by a constant vertex. In the derivation of \eq{mom5} we used condition \eq{mom3}, in this limit one can neglect
momentum transferred along Pomerons resulting into a constant vertex.
The action \eq{mom6} looks similar to the action of the toy model \eq{toy_S}. Following the procedure presented in
Section \ref{sec:GenFPI} the action of \eq{mom6} can be easily transformed to Langevin equation for directed percolation
\begin{eqnarray} \label{mom7}
 \frac{\partial \Phi(k,b;Y)}{\partial Y}
  =\frac{\bar{\alpha}^{}_S}{2\pi}\int d^2k^{\prime} K(k,k^{\prime})\Phi(k^{\prime},b;Y)
 -
  \frac{2\pi \bar{\alpha}^2_S}{N_c}\Phi(k,b;Y)\Phi(k,b;Y)+\zeta(k,b;Y)
\end{eqnarray}
with
\begin{eqnarray} \label{mom8}
<\zeta(k,b;Y)>=0
\end{eqnarray}
and
\begin{eqnarray} \label{mom9}
<\zeta(k,b;Y)\zeta(k^{\prime},b^{\prime};Y)>=\frac{4\pi \bar{\alpha}^2_S}{N_c}\;\Phi(k,b;Y)\;
\delta^{(2)}(\vec{b}-\vec{b^{\prime}})\;
\delta^{(2)}(\vec{k}-\vec{k^{\prime}})\;
\delta(Y-Y^{\prime})
\end{eqnarray}

The Langevin
equation is the one of many ways to describe a diffusion process and the considerable progress has been
achieved in this approach (see \cite{HH,WE,BIW,IT,MSW,SO}).

 However, we prefer a different way for description of the BFKL Pomeron interactions,  which will also  lead to diffusion equation: the so called generating functional approach. The advantage of the generating functional approach is its transparent relation to the partonic wave function of the fast hadron (dipole). In this approach we see
in the most explicit way our main theoretical problem: the BFKL Pomeron calculus provides the amplitude that
satisfies the $t$-channel unitarity while the $s$-channel unitarity is still a problem in the BFKL Pomeron
calculus. However, the probabilistic interpretation in the framework of the generating functional leads to the
correctly normalized partonic wave function which takes into account the main properties of the $s$-channel
unitarity as well.

\subsection{Generating functional: general approach}
In this subsection we discuss the main equations of the BFKL Pomeron Calculus in the
formalism of the generating functional,  which we consider as the most
appropriate technique for the probabilistic interpretation of this approach to
high energy scattering in QCD.

To begin with let us write down the definition of the generating functional
\cite{MUCD}
\begin{equation} \label{Z}
Z\left(Y\,-\,Y_0;\,[u] \right)\,\,\equiv\,\,
\end{equation}
$$
\equiv\,\,\sum_{n=1}\,\int\,\,
P_n\left(Y\,-\,Y_0;\,x_1, y_1; \dots ; x_i, y_i; \dots ;x_n, y_n
 \right) \,\,
\prod^{n}_{i=1}\,u(x_i, y_i) \,d^2\,x_i\,d^2\,y_i
$$
where $u(x_i, y_i) \equiv u_i $ is an arbitrary function of $x_i$ and $y_i$.
The coordinates $(x_i,y_i)$ describe the colourless pair of gluons or a dipole.
$P_n$ is a probability density to find $n$ dipoles with the size $x_i - y_i$
and  impact
parameter $(x_i + y_i)/2$.
It follows directly from the physical meaning of $P_n$ and the definition in \eq{Z} that the functional obeys the condition

\begin{equation} \label{ZIN1}
Z\left(Y\,-\,Y_0;\,[u=1]\right)\,\,=\,\,1\,.
\end{equation}
The physical meaning of this equation  is that the sum over
all probabilities is equal to unity.

Introducing vertices for the dipole process: $1 \to 2$ ($V_{1 \to 2}( x,y \to
x_1,y_1 + x_2,y_2)$) and   $2 \to 1$ ($V_{2 \to 1}( x_1,y_1 + x_2,y_2 \to x,y)$)
  we can write a typical birth-death equation in the form
\begin{equation}  \label{P}
\frac{\partial \,P_n(Y;\dots;x_i,y_i; \dots;x_n,y_n)}{\partial Y}\,\,\,=
\end{equation}
\begin{eqnarray}
 &=&\,\, \sum_{i} \  V_{1 \to
2}
\bigotimes \left(P_{n-1}(Y; \dots;x_i,y_i; \dots;x_n,y_n)\, -\,P_n (Y;\dots;x_i,y_i;
\dots;x_n,y_n)\right) \label{P1} \\
 &+ & \,\, \sum_{i > k} V_{2 \to 1}\,\bigotimes\,\left( P_{n+1}(Y; \dots;x_i,y_i; \dots ; x_k,y_k;\dots;x_n,y_n )
\,- \, P_n(Y;\dots;x_i,y_i;\dots ; x_k,y_k;\dots;x_n,y_n) \right)  \label{P2}
\end{eqnarray}
\eq{P} is the typical Markov chain and the fact that we have the correct normalized partonic wave function
is written in \eq{P} by introducing for each microscopic (dipole) process two terms (see \eq{P1} and  \eq{P2}): the emission of dipoles (positive birth term) and their  recombination (negative death term).
 Multiplying this equation  by the product $\prod^n_{i=1}\,u_i$
and integrating over all $x_i$ and $y_i$,  we obtain the
following linear equation for the generating functional:
\begin{equation}\label{ZEQ}
\frac{\partial \,Z\,\Lb Y-Y_0; [\,u\,]\Rb}{
\partial \,Y}\,\,= \,\,\chi\,[\,u\,]\,\,Z\,\Lb Y- Y_0; [\,u\,] \Rb
\end{equation}
with
\begin{eqnarray}
\chi[u]\,\,&=&\,\, \,\int\,d^4\,q d^4 q_1\,d^4 q_2 \,\,
  \left(  V_{1\,\rightarrow \,2}\left( q \to
q_1 + q_2 \right)\,  \left(  - u(q) \,+\,u(q_1) \,u(q_2)\,\right)
\,\frac{\delta}{\delta u(q)}\, \right. \label{chi} \\
 & &\left. - V_{2\,\rightarrow \,1}\left(  q_1 + q_2 \to q  \right)\,
\left( u(q_1)  \,u(q_2) \,-\, u(q) \right) \,\,\frac{1}{2} \,\frac{\delta^2}{\delta
u(q_1)\,\delta
u(q_2)} \right)\,
\label{VE21}
\end{eqnarray}
These equations describe the process of splitting of one Pomeron into two Pomerons and
 merging of two Pomerons into one Pomeron.  We will prove below, that this set of equations is
equivalent to the BFKL Pomeron calculus,  given by the path integral of \eq{BFKLFI}.
 We use notations $q_i$ for $(x_i,y_i)$ and $d^4 q_i $ for $d^2 x_i\,d^2\,y_i$
where $x_i$ and $y_i$ are positions of quark (antiquark) of the colourless dipole.

Eq.~(\ref{ZEQ})  is a typical diffusion equation or Fokker-Planck equation
\cite{GARD},   with the
diffusion
coefficient which depends on $u$. This is the master equation of our approach, and
the goal of this
paper is  to find the correspondence between this equation and  the BFKL Pomeron Calculus.
 In spite of the fact that this is a
functional equation we intuitively feel that this equation could be useful since
we can develop a
direct method for its  solution  and, on the other hand, there exist many  studies
of such
an equation
in the literature ( see for example  \cite{GARD}) as well as some physical
realizations in
statistical physics. The intimate relation between the Fokker-Planck equation, and
the high energy
asymptotic was first pointed out  by Weigert \cite{WE} in JIMWLK approach
\cite{JIMWLK},  and has
been discussed in
\cite{BIW,IT,MSW}.

The scattering amplitude can be defined as a functional \cite{K,L2}
\begin{eqnarray}
 N\left(Y;[\gamma_i] \right)\,&= & \,- \sum^{\infty}_{n =1} \int
(-1)^n\gamma_n(x_1,y_1;\dots;x_n,y_n;Y_0) \prod^n_{i=1}\frac{\delta}{\delta
u_i}Z\left(Y,[u_i]\right)|_{u_i=1}\,d^2 x_i\,d^2 y_i \label{N}\\
 &=& \,-\,\sum^{\infty}_{n =1} (-1)^n \int
\gamma_n(x_1,y_1;\dots;x_n,y_n;Y_0)\,\,\rho(x_1,y_1;\dots;x_n,y_n;Y -Y_0)\,d^2 x_i\,d^2 y_i \nonumber
\end{eqnarray}

The physical meaning of functions $\gamma_n$ is the imaginary part of the amplitude
of interaction of $n$-dipoles with the target at low energies. All these functions
should be taken from the non-pertubative QCD input. However, in
\cite{L1,L2,L3}
it was shown that we can introduce the amplitude of interaction of $n$-dipoles
$\gamma_n(x_1,y_1;\dots;x_n,y_n;Y)$ at
high energies (large values of rapidity $Y$) and  \eq{Z}, \eq{ZEQ}
and \eq{N} can be rewritten as a chain set of equation for
$\gamma_n(x_1,y_1;\dots;x_n,y_n;Y)$.  The equation has the form\footnote{This
equation is Eq.~(2.19) in  \cite{L3} but, hopefully, without  misprints,
part of which has been noticed in \cite{IST}.}

\begin{eqnarray} \label{N5}
{}{}  \frac{\partial\, \gamma_n\left( q_1\dots, q_n
\right)}{\partial\,Y}&=&  2 \,\sum_{i=1}^n\,\int\,d^4q'\,d^4q \,
V_{1\,\rightarrow \,2} (q_i;\,q,\,q')\,
\gamma_n\left(  \dots q'\dots \right) \,\nonumber
-\,   \sum_{i=1}^n \,\int d^4q_1'\,d^4q_2'\,
V_{1 \,\rightarrow\,2} (q_i;\,q_1',\,q_2')\,\gamma_n\left(  \dots, q_i\dots
\right)
 \\
&-&\sum^{n - 1}_{i =1} \int d^4q\,d^4q'V_{1\,\rightarrow \,2} (q_i;\,q,\,q')\,\gamma_{n+1}\left(  \dots q \dots
q'\right)
-\,\sum_{i > j}^n\,\int  d^4\,q\,
V_{2\,\rightarrow\,1}\left( q_i,\,q_j ;\,q \right) \gamma_{n-1}\left(
 q_i \dots  q_j \dots q\right) \nonumber
\\
&+&\, 2\,\sum_{i=1}^n \int d^4 q\, d^4 q' V_{2
\rightarrow 1} \left( q ,q_i;q'\right) \, \gamma_{n-1}\left( \dots q_i \dots
q\right)
\, +
\,\sum^n_{i > k}\,   \int d^4q\,
V_{2\rightarrow 1} \left( q_i ,q_k; q \right)\,\gamma_n\left(  \dots
q_i\dots q_k \dots \right)
\nonumber\\
\end{eqnarray}

Comparing this equation for $\gamma_1 \equiv T^{(1)}$ and $\gamma_2 \equiv T^{(2)}$
one can see that
\begin{eqnarray}
V_{1 \to 2} &=&\frac{\bar{\alpha}_S}{2\,\pi} \Gamma_{1 \to
2}\,\,=\,\,\frac{\bar{\alpha}_S}{2\,\pi}\,K\left(x,y;z\right)\,; \label{GA1}\\
V_{2 \to 1}\,&=&\, \,\frac{2\,\alpha^2_S}{\pi^2}
\frac{\bar{\alpha}_S}{2 \,\pi }\, \left(-\,\frac{\Gamma_{2 \to
1}(x_1,y_1 + x_2,y_2 \to x,y)}{(x - y)^4} \,+ \right.
 \label{GA2}\\
&+& \left.
 \,
\,\int \,\frac{d^2x\,d^2y}{(x - y)^4}\,\Gamma_{2 \to 1}(x_1,y_1 +
x_2,y_2 \to x,y) \,
\left(\delta^{(2)}(x_1 - x)\delta^{(2)}(y_1 - y)\,+\,\delta^{(2)}(x_2 -
x)\delta^{(2)}(y_2
- y)\right) \right)\nonumber
\end{eqnarray}
with $\Gamma_{2 \to 1}$ is given by \eq{V21}.
\subsection{A toy model: Pomeron interaction and probabilistic interpretation}
In this section we consider the simple toy model in which the probabilities to find
$n$-dipoles being  independent  on  dipoles sizes \cite{MUCD,L1,L3,L4}. In this
model the master equation (\ref{ZEQ}) has a simple form
\begin{equation} \label{TMZ}
\frac{\partial Z}{\partial Y}\,=\,- \Gamma(1 \to 2)\,u (1 - u)
\,\frac{\partial Z}{\partial u}+\,\frac{1}{2}\,\Gamma(2 \to
1)\,u (1 - u)\,\frac{\partial^2 Z}{\partial u^2}
\end{equation}
\eq{TMZ} generates: the Pomeron splitting $G_{P \to 2P}=  \Gamma(1 \to
2)$;  Pomerons merging $G_{2P \to P}=  \frac{1}{2}\Gamma(2 \to 1)$  and also the two Pomerons
scattering $G_{2P \to 2P}\,\,= \, \frac{1}{2}\Gamma(2 \to 1)$.  It is easy to see that by
neglecting $u^2 \partial^2 Z/\partial u^2$ term in Eq.~(\ref{TMZ}) we cannot
provide a correct sign for Pomerons merging $G_{2P \to P}$. It worth mentioning that we need this term to provide the conservation of the total probability on each level of rapidity.

The description given by \eq{TMZ} is equivalent to the path integral of \eq{toy_S}. To see this we need
to notice that the general solution of \eq{TMZ} has a form
\beq \label{TMH1}
Z(Y;u)\,\,=\,\,e^{H(u)\,(Y - Y_0)}Z(Y_0;u)
\eeq
with operator $H$ defined as
\beq \label{TMH2}
H(u)\,\,=\,\,-\, \Gamma(1 \to 2)\,u (1 - u)
\,\frac{\partial }{\partial u}+\, \frac{1}{2}\Gamma(2 \to
1)\,u (1 - u)\,\frac{\partial^2 }{\partial u^2}
\eeq
and
\beq \label{TMZ0}
Z(Y_0;u)\,\,=\,\,e^{\tau_{tr}( u -1 )}
\eeq

Introducing operators of creation ($a^+$) and annihilation ($a$)
\beq \label{TMH3}
\hat{a}\, =\,\frac{\D}{\D u}\,; \,\,\,\,\,\,\,\,\,\, \hat{a}^+ \,=\,u \,\,\,\mbox{that satisfy} \,\,[
\hat{a} ,\hat{a}^+] = 1
\,\,\, \mbox{at fixed}\,\, Y
\eeq
one can see that operator ${\cal H}$ has the form
\beq \label{TMH4}
{\cal H}\,\,=\,\,-\,\Gamma(1 \to 2)\,\hat{a}^+\,( 1 - \hat{a}^+)\,\hat{a} \,\,+\,\,\frac{1}{2}\Gamma(2 \to
1)\,\hat{a}^+\,( 1 - \hat{a}^+)\,
\hat{a}^2
\eeq
and the initial state at $Y=Y_0$ is defined as
\beq \label{TMIS}
|Y_0>\,\,=\,\,e^{\tau_{tr}( \hat{a}^+ \,-\,1)} |0>
\eeq
with the vacuum defined by $\hat{a} |0> = 0$.

We need to discretize the development operator of \eq{TMH1} with ${\cal H}$ given by \eq{TMH4},
namely,
\beq \label{TMH5}
e^{{\cal H}\,(Y - Y_0)}=e^{{\cal H}\,\Delta Y}\, \dots\, e^{{\cal H}\,\Delta Y}
=\,\prod^N_{j=1}\,\Lb 1 \,+\,{\cal H}\,\Delta Y  \Rb
\eeq

and introduce   coherent states \cite{CION}
for a certain intermediate rapidity $Y_j=Y_0+j\Delta Y$
 as
\beq \label{TMH6}
|\phi_j>\,\,=\,\,e^{\phi_j\,\hat{a}^+ \,-\,\phi_j }\,|0>
\eeq
where $\phi_j$ are arbitrary complex numbers.
The initial state of \eq{TMIS} can be written as
\begin{center}
\begin{eqnarray}
|\phi_0(Y_0)>\,\equiv| \tau_{tr}>.
\end{eqnarray}
\end{center}

The unit operator in terms of the coherent states can be expressed as

\beq \label{TMH7}
1\,=\,\Lb \,\int\,\frac{d \phi_j d \phi_j^*}{\pi i}\,e^{ - \phi_j\,\phi_j^*\,+\,\phi_j\,+\,\phi_j^*} \Rb
|\phi_j>\,<
\phi_j|
\eeq

We want to calculate matrix element of some operator $A$ between states of initial $Y_0$ and
 final $Y$ rapidity $<Y|A|Y_0>$. This can be written as

\beq
<Y|A|Y_0>=<Y|A\left\{
\,\Lb \,\int\,\frac{d \phi_Y d \phi_Y^*}{\pi i}\,e^{ - \phi_Y\,\phi_Y^*\,+\,\phi_Y\,+\,\phi_Y^*} \Rb
|\phi_Y>\,<
\phi_Y|
\right\}|Y_0>
\eeq
here we denote $|Y>\equiv |\phi_Y>$.
Next we use the development operator given in \eq{TMH5} to find $<Y|Y_0>$. We split
the rapidity $Y-Y_0$ to $N$ intervals and insert the development \eq{TMH5} and unit \eq{TMH7}
 operator  between the states of intermediate rapidity
\beq
<Y|   \prod^{N}_{j=1}\,\Lb 1 \,+\,{\cal H}\,\Delta Y  \Rb\, |Y_0>
\eeq
 We look at
\begin{eqnarray}
<\phi_{j+1}|\Lb 1 \,+\,{\cal H}\,\Delta Y
\Rb|\phi_j>&=&\exp\left\{-\phi^*_{j+1}\phi_{j+1}+\phi^*_{j+1}+\phi_{j+1}
-\phi^*_{j+1}-\phi_{j}+\phi^*_{j+1}\phi_j \right\}\, \nonumber \\
&\times&
\Lb 1 \,+ \,{\mathcal{H}( \phi^*_{j+1}, \phi_j
)}\,\Delta Y
 \Rb \nonumber \\
 & =&
\exp\left\{-\phi^*_{j+1}(\phi_{j+1}-\phi_{j})+\phi_{j+1}-\phi_{j}\right\}\Lb 1 \,+ \,{\mathcal{H}(
\phi^*_{j+1}, \phi_j )}\,
\Delta Y
  \Rb \nonumber \\
&= & \exp \left\{-\phi^*_{j+1}(\phi_{j+1}-\phi_{j})+\phi_{j+1}-\phi_{j}\right\}\,\exp\left(\mathcal{H}(
\phi^*_{j+1}, \phi_j )
\,\Delta Y\right)  \label{alex1}
\end{eqnarray}

Now we redefine an arbitrary function $\phi_j$ as
\beq
\Phi^{+}_j\,\,\,=\,\,\,-\phi_j,   \hspace{2cm} \Phi_{j}=1-\phi^*_{j}
\eeq
and rewrite \eq{alex1} in terms of $\Phi_j$ and $\Phi^+_j$
\begin{eqnarray}
\,\,\,\,\,\,\,\,\,\,\,\,& &
e^{-\phi^*_{j+1}(\phi_{j+1}-\phi_{j})+\phi_{j+1}-\phi_{j}}\,\,\exp\left\{\mathcal{H}( \phi^*_{j+1},
\phi_j
)\,\Delta Y\right\}\,\,= \nonumber \\
\,\,\,\,\,\,\,\,\,\,\,\, &=&
 \exp\left\{\Phi_{j+1}(\Phi^{+}_{j+1}-\Phi^{+}_{j})+\mathcal{H}( 1-\Phi_{j+1}, -\Phi^{+}_j
)\right\}\,\label{alex2} \\
& =& \exp\left\{  \left( \frac{\Phi_{j+1}(\Phi^{+}_{j+1}-\Phi^{+}_{j})}{\Delta Y}+\mathcal{H}( 1-\Phi_{j+1},
-\Phi^{+}_j
 )\,  \right )\Delta Y  \right\} \nonumber
\end{eqnarray}
 Summing over all rapidity intervals we have
\beq
<Y|A|Y_0>\,\,\,
\sim\,\,\,\prod_{j=0}^{N}\int d\Phi_j^+ d\Phi_j A(Y) e^{S}
\eeq
where $A(Y)$ is the expectation value of the operator $A$ at the final rapidity $Y$, and
\beq
S\,\,=\,\,\left( \frac{\Phi_{j+1}(\Phi^{+}_{j+1}-\Phi^{+}_{j})}{\Delta Y}+\mathcal{H}( 1-\Phi_{j+1}, -\Phi^{+}_j
)\,  \right )\Delta Y
\eeq

In the continuous limit this becomes
\beq \label{FUIN}
<Y|A|Y_0>  \,\,\,=\,\,\,\frac{ \int \mathcal{D}\Phi^+ \mathcal{D}\Phi \,\,A(Y)\,\, e^{S}}{\int \mathcal{D}\Phi^+
\mathcal{D} \Phi\,\, e^{S}}
\eeq
with
\begin{eqnarray}\label{TMHAC}
  S\,&=&\,\int dY \left( \Phi^+  \frac{d}{dY} \Phi +\mathcal{H} ( \Phi +1, -\Phi^+ ) \, \right)
 \,\, \\
&=&\, \,\int dY   \left( \Phi^+  \frac{d}{dY} \Phi \,
-\Gamma(1\rightarrow2) \Phi^+\Phi
+\Gamma(1\rightarrow2) \Phi^{+}\Phi^2
+\frac{1}{2}\Gamma(2\rightarrow 1) \Phi^{+2}\Phi
-\frac{1}{2}\Gamma(2\rightarrow 1) \Phi^{+2}\,\Phi^2
 \, \right)  \nonumber
\end{eqnarray}

\eq{TMHAC} is  the action of \eq{toy_S} for $\Gamma(1 \to 2) = \frac{1}{2}\Gamma(2 \to 1)$. In the
toy-model the difference between these two vertices is the normalization problem of functions
$\Phi^+$ and $\Phi$. In our approach they are normalized in the way which allows us to treat them as
probabilities (see \eq{T} and \eq{TN}).  However, \eq{TMHAC} includes the new interaction: a
transition of two Pomerons to two Pomerons. The sign is such that this interaction provides the
stability of the potential energy. Indeed this term is responsible for the increase of the potential
energy at large values of both $\Phi^+$ and $\Phi$.

Comparing \eq{TMHAC} with \eq{SM1} one can see that we built the partition function and the
thermodynamic potential using the generating functional. It means that our \eq{ZEQ} is equivalent to
statistical description of the system of dipoles.

Eq.~(\ref{TMZ}) is the diffusion with the $u$ dependence in  diffusion coefficient.
In terms of the Langevin equation \eq{TMHAC} generates a noise term
of the $\;\;<\zeta \zeta> \propto \Phi(1-\Phi) \;\;$ type.

 To out taste \eq{TMZ} is simpler than the Langevin equation of \eq{noise7} and it will be easily
generalized for the case of QCD.     For $u<1$ the diffusion coefficient is positive and the equation
has a reasonable solution. If $u >1$, the sign of this coefficient changes and the equation gives a
solution which increases with $Y$ and $Z(Y)$  cannot be treated as the
generating function for the probabilities to find $n$ dipoles (Pomerons) (see
\cite{GRPO,BOPO,L4} for details). The same features one can notice in the asymptotic
solution that is the solution to Eq.~(\ref{TMZ}) with the l.h.c. equal to zero.
It is easy to see that this solution has the form
\begin{equation}
Z(Y \to \infty; u ) \,\,=\,\,\frac{1 \,-\,e^{\,\kappa\, u}}{1\,-\,e^{\,\,\kappa}}\,;\,\,\,\mbox{with}\,\,\,
\kappa\,=\,
2\frac{\Gamma(1 \to 2)}{\Gamma(2 \to 1)}\,=2\,\frac{2
\,N^2_c}{\bar{\alpha}^2_S}\,\gg\,1
\end{equation}
For negative $\kappa$ this solution leads to $Z>1$ for $u <1$.
This shows that we cannot give a probabilistic interpretation for such a solution.

\section{A practical way to find solution: Monte Carlo simulation}

In this section we consider the BFKL Pomeron calculus in the form of \eq{mom5} which leads to the simplest
approach in the framework of the generating functional technique. This approach is not only the simplest one  but
also it is free from all troubles related to the negative contribution for the process of $2P \to P$
transition. Repeating procedure discussed in section 3.2 one can see that  \eq{mom5} for action leads to the
following equation for the generating functional
\begin{equation}\label{ZEQMR}
\frac{\partial \,Z\,\Lb Y-Y_0; [\,u(b,k)\,]\Rb}{
\partial \,Y}\,\,= \,\,\chi\,[\,u(b,k)\,]\,\,Z\,\Lb Y- Y_0; [\,u(b,k)\,] \Rb
\end{equation}
with
\begin{eqnarray}
\chi[u]\,\,&=&\,\, \,\int\,d^2\,b\, d^2\, k\, \left( \frac{\bas}{2 \pi}
 \left(- \int\,d^2\,k'\, K(k,k') u(b,k')\,\frac{\delta}{\delta u(b,k')} \,+\,u(b,k) \,u(b,k)\,
\,\frac{\delta}{\delta u(b,k)} \right)\right. \label{chimr} \\
 & &\left. - \left( \frac{4\,\pi^2 \bas}{N_c} \right)^2\,\frac{\bas}{2 \pi}
\left( u(b,k)  \,u(b,k) \,-\, u(b,k) \right) \,\,\frac{1}{2} \,\frac{\delta^2}{\delta
u(b,k)\,\delta
u(b,k)} \right)\,
\label{VE21MR}
\end{eqnarray}
For  $Z(Y,[u(b,k)])$ defined as \beq \label{ZMR}
Z\left(Y\,-\,Y_0;\,[u(b_i,k_i)] \right)\,\,\equiv\,\,
\eeq
$$
\equiv\,\,\sum_{n=1}\,\int\,\,
P_n\left(Y\,-\,Y_0;\,b_1, k_1; \dots ; b_i, k_i; \dots ;b_n, k_n
 \right) \,\,
\prod^{n}_{i=1}\,u(b_i, k_i) \,d^2\,x_i\,d^2\,k_i
$$
where $u(b_i,k_i)$ are  arbitrary functions.

\eq{chimr} and \eq{VE21MR} show that the $2P \to 1P$ transition can be written as the two dipole to one dipole
merging with a positive probability. Therefore, \eq{ZEQMR} has a very simple probabilistic interpretation
which can be written as the following Markov chain:
\begin{equation}  \label{PMR}
\frac{\partial \,P_n(Y;\dots;b_i=b,k_i=k; \dots;b_n=b,k_n=k)}{\partial Y}\,\,\,=
\end{equation}
\begin{eqnarray}
 &=&\,\,  \frac{\bas}{2 \pi}\,\sum_{i} \left(P_{n-1}(Y; \dots;b_i=b,k_i=k; \dots;b_{n-1},k_{n
-1})\,- \right.\nonumber\\
 & &\left. -\,\,\int \,d^2\,k' \,K(k,k')\,\,P_n
(Y;\dots; b_i=b,k_i=k';
\dots;x_n=b,k_n=k)\right) \label{PMR1} \\
 &+ & \,\, \left( \frac{4\,\pi^2 \bas}{N_c} \right)^2\,\frac{\bas}{2 \pi} \sum_{i > j} \left( P_{n+1}(Y;
\dots;b_j=b,k_j=k; \dots ;
b_i=b,k_i=k;\dots; b_n=b,k_n=k ) -\right.\nonumber\\
 & &\left.\,- \, P^{}_n(Y;\dots;b_j=b,k_j=k;\dots b_i=b,k_i=k; \dots; b_n=b,k_n=k;) \right);  \label{PMR2}
\end{eqnarray}

This set of equations can be solved numerically and it gives a practical way to discuss the influence
of the Pomeron loops on the solution for the scattering amplitude at high energies.

\section{Conclusions}
We demonstrate in this paper that the BFKL Pomeron Calculus in the kinematic region given
by \eq{REPC} has two equivalent descriptions: (i)\,\,one is the generating functional  which  gives a  clear
probabilistic
interpretation of the processes of high energy scattering and   provides also a Hamiltonian-like
description of the
system
of interacting dipoles; (ii)\,\,the second is the Langevin equation with a noise term of \eq{noise4}
 which is rather complicated.
We show that at high energies the Langevin equation with noise of \eq{noise4} can be reduced to  the Langevin equation for directed
percolation in the momentum representation if the impact parameter is large, namely,
$b \,\gg 1/k$ where $k$ is the transverse momentum of a dipole. Unfortunately, this simplified form of
Langevin equation is not applicable for summation of Pomeron loops, where one integrates over all
possible values of impact parameter.

 In other words, the BFKL Pomeron Calculus can be considered as an alternative description
of the statistical system of dipoles with different kinds of interactions between them. The same conclusion
was drawn in  \cite{BON} using a similar technique.
We show that the BFKL Pomeron calculus with two vertices: splitting of one Pomeron into two Pomerons($P \to 2\,P$) and
merging of two Pomerons into one Pomeron ($2 P \,\to\,P$), can be described as a system of colourless dipoles
with two processes: the decay of one dipole into two dipoles and the merging of two dipoles into one dipole.

It is shown that the question about negative amplitude does not arise  if we treat the system of dipoles
 in the momentum representation. Markov chain  for this system is written in the paper (see \eq{PMR})
   and can be considered as a practical way to find a solution in accessible range of energies.
However, as we have stressed a number of assumptions we have made on the way to
 simplify the noise term as well as to argue the probabilistic interpretation of
 the Pomeron calculus. Therefore, both  these approaches can be considered as a QCD motivated models.
 In the case of the probabilistic interpretation we can even expect that
this approach  will not work in the form, that has been suggested in this paper,
  for the full ($2 P \,\to\,P$)\ vertex (see  \cite{IST}).

 Being elegant and beautiful the BFKL
Pomeron Calculus has a clear
disadvantage: it  lacks   theoretical ideas what kind of Pomeron interactions we
should take into account and why.  Of course,  Feynman diagrams in leading
$\ln(1/x)$ approximation of perturbative QCD allow us, in principle, to calculate all
possible Pomeron
interactions but, practically, it is very hard job. Even if we  calculate these
vertices we need to understand what set of vertices we should take into account for
 the calculation of the scattering amplitude.
 This is the reason why we need to
develop a more general formalism. Fortunately, such a formalism has been
built and it is known under the abbreviation JIMWLK-Balitsky approach
\cite{MV,JIMWLK,B}. In this approach we are able to calculate all vertices for
Pomeron interactions as it was demonstrated in  \cite{KOLU} and it solves the
first part of the problem: determination of all possible Pomeron interactions.
However, we need to understand what vertices we should take into account for
calculation of the scattering amplitude. We hope that a further progress in going
beyond of the BFKL Pomeron Calculus (see \cite{KOLU,HIMS}) will lead to such
a development of the BFKL Pomeron Calculus with  a consistent
theoretical
approach. Hopefully this approach will be simpler than Lipatov effective action
\cite{LIEF} which is not easier to solve than the full QCD Lagrangian.

It is well known that the mean field approach to our problem,  which
includes only one dipole  to two dipoles decay in the master
equation (see \eq{ZEQ}) has been studied quite well both
analytically \cite{LT} and numerically \cite{THEORVRSDATA}. We
firmly believe that  the probabilistic interpretation in spite of
being a QCD motivated model, gives a practical method for creating a
Monte Carlo code in spirit of the approach suggested in \cite{MS}.
This code will allow us to find a numerical solution to the problem
and to consider   inclusive observables. This extension is very
desirable since the most experimental data exist for these
observables.

\section*{Acknowledgments}
We want to thank Asher Gotsman, Larry McLerran, Dima Kharzeev, Alex Kovner, Misha Lublinsky and Uri
Maor for very useful
discussions on the subject
of this paper. Our special thanks go to Jose Guilherme Milhano, who draw our attention to  \cite{HH,DMS}
and the discussions with whom on the subject of this paper were very useful and instructive.

 This research was supported in part  by the Israel Science Foundation,
founded by the Israeli Academy of Science and Humanities and by BSF grant \# 20004019.
\appendix
\begin{boldmath}
\section{ Calculation of $G_0(x_1,x_2|x'_1,x'_2)$}  \label{sec:A}
\end{boldmath}
The solution of the BFKL equation is given by \cite{LI}
 
 \beq \label{sol_pr}
G(x_1,x_2;x^{'}_1,x^{'}_2|\omega)\,=\, 
\sum_{n=-\infty}^{+\infty}\int_{-\infty}^{+\infty}
\frac{(\nu^2+n^2/4) d\nu}{[\nu^2+(n-1)^2/4][\nu^2+(n+1)^2/4]}
\frac{G_{\nu \mu}(x_1,x_2,x^{'}_1,x^{'}_2)}{\omega-\omega(\nu, \mu)}
 \eeq

where $G_{\nu \mu}(x_1,x_2,x^{'}_1,x^{'}_2)$ is the Mellin transform of 
\eq{BFKLGF} and
$x_i$ are two-dimensional vectors in complex coordinates 
  \begin{eqnarray}\label{xcomp_pr}  
  x_i=x_{i,x}+i x_{i,y} & &  x^*_i=x_{i,x}-i x_{i,y}
  \end{eqnarray} 
The function $\omega(\nu, \mu)$ is the eigen value of the BFKL equation
given by  \eq{BFKLOM}.

The four-point Green function is presented in terms of the hypergeometric 
functions \cite{LI,NP}
  \begin{eqnarray}\label{4Green_pr}
  &  G_{\nu \mu}(x_1,x_2,x^{'}_1,x^{'}_2)=C_1 x^h x^{*\tilde{h}}
F(h,h,2h;x) F(\tilde{h},\tilde{h},2\tilde{h};x^*) & \nonumber \\
&+C_2 x^{1-h} x^{* 1- \tilde{h}}
F(1-h,1-h,2-2h;x) F(1-\tilde{h},1-\tilde{h},2-2\tilde{h};x^*) &
  \end{eqnarray}
with $h=\frac{1}{2}+i \nu +\frac{n}{2}$, $\tilde{h}=\frac{1}{2}+i \nu
-\frac{n}{2}$, and $x$ is the anharmonic ratio 
  
  \beq \label{anhar_rat}
    x= \frac{x_{12} x_{1' 2'}}{x_{1 1'} x_{22'}}
  \eeq
Coefficients  $C_1$ and $C_2$ are given by \cite{LI} 
  \begin{eqnarray}\label{C1C2_pr}
    C_1=\frac{b_{n,-\nu}}{2 \pi^2} \hspace{2cm} 
    C_2=\frac{b_{n,\nu}}{2 \pi^2}  
  \end{eqnarray}
with 
  \begin{eqnarray}\label{bnu_pr}
  b_{n,\nu}=\pi^3 2^{4i\nu}
\frac{\Gamma(-i\nu+(1+|n|)/2) \Gamma(i\nu+|n|/2)}{\Gamma(i\nu
+(1+|n|)/2)\Gamma(-i\nu+|n|/2)}
  \end{eqnarray}
As we have discussed, the high energy asymptotic behaviour stems from $n=0$ term in  
\eq{sol_pr}. The initial condition for \eq{BFKLGF} at $Y=Y_0$  is given by the 
following expression
\beq\label{sol1_pr}
G_0(x_1,x_2;x'_1,x'_2)\,\,=
\eeq
$$
\,\,\int\,\frac{d \omega}{2 \,\pi 
\,i} \int_{-\infty}^{+\infty}\,\,
\frac{\nu^2 d\nu}{(\nu^2+1/4)^2}
\frac{G_{\nu \mu}(x_1,x_2,x'_1,x'_2)}{\omega - \omega(\nu,n=0)}\,=\,\frac{1}{\omega}
\int_{-\infty}^{+\infty}\,\,
\frac{\nu^2 d\nu}{(\nu^2+1/4)^2}\,\,G_{\nu \mu}(x_1,x_2,x'_1,x'_2)
$$
This integral can be taken by closing contour of integration over
singularities of the integrand.

The function 
\begin{eqnarray}\label{nu2_pr}
\frac{\nu^2 }{(\nu^2+1/4)^2}
\end{eqnarray}
has two poles at $\frac{i}{2}$ and $-\frac{i}{2}$.
The four-point Green function of  \eq{4Green_pr} consists of two terms, one
with $C_1(x x^*)^{\frac{1}{2}+i\nu}$ and the other one $C_2(x
x^*)^{\frac{1}{2}-i\nu}$.
 For small $|x|$ this terms should be integrated
closing contour in  upper and lower semi-planes respectively. 
The resulting contour in the lower  semi-plane runs
 anticlockwise
and thus the value of the contour integral enters with a minus sign.

The terms
could be expanded in the vicinity of their poles. Let us consider the
first term. We expand the function $C_1$ in the vicinity of $\frac{i}{2}$
  \begin{eqnarray}\label{C1exp_pr}
  C_1 = \frac{\pi}{2}2^{-4i\nu}
\frac{\Gamma(-i\nu)}{\Gamma(-i\nu+\frac{1}{2})\Gamma(1+i\nu)} 
   \frac{\Gamma(i\nu+\frac{3}{2})}{(i\nu+\frac{1}{2})}
  \end{eqnarray}
The hypergeometric function can be written as a sum
  \begin{eqnarray}\label{hypersum_pr}
   F(a,b,c;x)=1+\frac{\Gamma(c)}{\Gamma(a)\Gamma(b)}\sum^{\infty}_{n=1}
   \frac{\Gamma(a+n)\Gamma(b+n)}{\Gamma(c+n)}\frac{x^n}{n!}
  \end{eqnarray} 
In the case of
$F(h,h,2h;x)$ the
singularity of $\Gamma$ function at $\frac{i}{2}$ can factorized out the
sum
  \begin{eqnarray}\label{hyperln_pr}
   F(h,h,2h;x)=1+\frac{\Gamma(2h)}{\Gamma(h)\Gamma(h)}\sum^{\infty}_{n=1}
   \frac{\Gamma(h+n)\Gamma(h+n)}{\Gamma(2h+n)}\frac{x^n}{n!}\nonumber\\
  \simeq 1+\frac{1}{2\Gamma(h)}\sum^{\infty}_{n=1}
   \frac{x^n}{n}=1-\frac{1}{2} \frac{ln(1-x)}{\Gamma (i\nu + \frac{1}{2} )}
  \end{eqnarray}
At this stage the first term of the integrand of \eq{sol1_pr} can be
 written as
   \beq\label{1term_pr}
\eeq
$$
\frac{1}{\omega}\frac{\nu^2}{(i\nu+\frac{1}{2})^3(i\nu-\frac{1}{2})^2} 
\frac{\pi}{2}  2^{-4i\nu} \frac{\Gamma(-i\nu)}{\Gamma(-i\nu+\frac{1}{2})\Gamma(1+i\nu)} 
 (xx^*)^{\frac{1}{2}+i\nu}
\left(1-\frac{1}{2}\frac{ln(1-x)}{\Gamma(\frac{1}{2}+i\nu)}\right)
\left(1-\frac{1}{2}\frac{ln(1-x^*)}{\Gamma(\frac{1}{2}+i\nu)}\right) 
 $$
 
It is clearly seen that the term of zero order in $ln|1-x|$  has a 
third order pole at $\nu=\frac{i}{2}$; the term of first order in $ln|1-x|$
has a second order pole at $\nu=\frac{i}{2}$, and, the term of second order in
$ln|1-x|$ has a simple pole at $\nu=\frac{i}{2}$. The contributions of those
terms in the contour integral are found to be 
   \beq\label{C1zero_pr}
   \frac{i \pi^2}{\omega}[-2\,\,+
\,\, ln(xx^*)(-4 + ln(xx^*) )]
   \eeq
   \beq\label{C1first_pr}
   -\frac{ \pi^2}{2\omega}
[-2\,\,+\,\,\gamma\,\, + \,\, ln(xx^*)]ln[(1-x)(1-x^*)]
   \eeq
    \beq\label{C1second_pr}
  - \frac{i\pi^2}{4\omega}ln(1-x)ln(1-x^*)
   \eeq
respectively.

In a similar way we may expand the second of the integrand in vicinity 
of its pole at $\nu=-\frac{i}{2}$, namely

\beq \label{2term_pr}
\eeq
$$
\frac{1}{\omega}\frac{\nu^2}{(\frac{1}{2}-i\nu)^3(i\nu+\frac{1}{2})^2} 
\frac{\pi}{2}  2^{4i\nu} \frac{\Gamma(i\nu)}{\Gamma(i\nu+\frac{1}{2})\Gamma(1-i\nu)} 
 (xx^*)^{\frac{1}{2}-i\nu}
\left(1-\frac{1}{2}\frac{ln(1-x)}{\Gamma(\frac{1}{2}-i\nu)}\right)
\left(1-\frac{1}{2}\frac{ln(1-x^*)}{\Gamma(\frac{1}{2}-i\nu)}\right) 
$$
  The integration is performed on the lower semicircle and  results in overall minus sign 
  of the integral. The contributions corresponding to \eq{C1zero_pr}, 
\eq{C1first_pr} 
and \eq{C1second_pr} are 

\beq\label{C2zero_pr}
-\frac{i \pi^2}{\omega}[-\,\,2\,\,+\,\,
 ln(xx^*)(-4 + ln(xx^*) )]
   \eeq
   \beq \label{C2first_pr}
   -\frac{ \pi^2}{2\omega}
[-2\,\,+\,\,\gamma \,\,+\,\,  ln(xx^*)]ln[(1-x)(1-x^*)]
   \eeq
\beq \label{C2second_pr}
  + \frac{i\pi^2}{4\omega}ln(1-x)ln(1-x^*)
   \eeq
   respectively.

Comparing the contributions we note that that of zero and second order \\ in $ln[(1-x)(1-x^*)]$ are 
exactly canceled out, and we are left with 
\beq \label{finalCont_pr}
   -2\frac{ \pi^2}{2\omega}
[-2\,\,+\,\,\gamma \,\,+\,\,  ln(xx^*)]ln[(1-x)(1-x^*)]
   \eeq 
   For small $|x|$ this can written as
   \beq \label{final2_pr}
   -\frac{4\pi^2}{\omega}\hspace{0.15cm}ln|x|\hspace{0.15cm}ln|1-x|
   \eeq 
   or
   \beq\label{final02_pr}
  +\frac{4\pi^2}{\omega}\hspace{0.15cm}ln\frac{1}{|x|}\hspace{0.15cm}ln|1-x|
   \eeq 
   Going  back to complex vector representation of x and rewriting \eq{final3_pr}  as
   \beq \label{final3_pr}
G_0(x_1,x_2;x'_1,x'_2)\,\,=\,\,
  +\frac{4\pi^2}{\omega}\hspace{0.15cm}ln \left|\frac{x_{11'}x_{22'}}{x_{12'}x_{1'2}}\right|  
   \hspace{0.15cm}ln \left|\frac{x_{11'}x_{22'}}{x_{12}x_{1'2'}}\right|
   \eeq
we see that we    reproduce the result of \cite{LI}. Therefore, we demonstrated that    
\eq{final3_pr} gives a correct initial condition for searching the scattering 
amplitude at high energies restricting ourselves by the one term in \eq{BFKLGF} with 
$n =0$. 

\section{ The path integral formalism for the generating  functional} \label{sec:B}
We want to develop a path integral formalism similar to that we
found for the toy model, but where the probabilities to find $n$
dipoles depend of dipole sizes. In our notation we denote by Latin
index rapidity interval, and Greek indices relate to a size of
dipole.

As in Chapter 3.3  we introduce the creation and annihilation operators
\begin{center}\label{appB_1}
\begin{eqnarray}
\hat{a}(q)=\frac{\delta}{\delta u(q)} \hspace{2cm} \hat{a}^{\dagger}(q)=u(q)
\end{eqnarray}
\end{center}
with commutation relations $[\hat{a}(q),\hat{a}^{\dagger}(q')]=\delta(q - q')$ at fixed $Y$. The expression for the coherent states in this case takes form of
\begin{center}
\begin{eqnarray}
|\phi(q)>=e^{\phi(q)\hat{a}^{\dagger}(q)-\phi(q)}|0>
\end{eqnarray}
\end{center}
with
\begin{center}
\begin{eqnarray} \label{appB_11}
\hat{a}(q')|\phi(q)>=\phi(q)|\phi(q)> \delta(q -q')
\end{eqnarray}
\end{center}

First we consider discrete dipole sizes $q_\alpha=(L/N)\alpha$, where $L$ is a maximal possible dipole size, $N$
a number of intervals of $L$ discretization, and $\alpha$ is an integer number running from $0$ to $N$.
In this case the commutation relations become $[\hat{a}(q_\alpha),\hat{a}^{\dagger}(q_{\beta})]=\delta _{\alpha,\beta}$.

The unit operator can be written in terms of the coherent states
\begin{center}
\begin{eqnarray}
\hat{I}=\prod_{\alpha}\int \frac{d\phi^*(q_\alpha) d\phi(q_\alpha)}{i\pi } e^{ -\phi^*(q_{\alpha})\phi(q_\alpha)+\phi(q_\alpha)+\phi^*(q_\alpha)}  |\phi(q_\alpha)><\phi(q_\alpha)|
\end{eqnarray}
\end{center}

The operator $\mathcal{H}$ defined in Eq.~(\ref{chi}) can be written as
\begin{center}
\begin{eqnarray} \label{appB_2}
\mathcal{H}[\hat{a}^{\dagger},\hat{a}] = &-& \sum_{\beta} \sum_{\gamma} \sum_{\lambda} [V_{1\rightarrow 2}
(q_\beta\rightarrow q_\gamma+q_\lambda) \{-\hat{a}^{\dagger}(q_{\beta}) +\hat{a}^{\dagger}(q_{\gamma})\hat{a}^{\dagger}(q_{\lambda}) \} \hat{a}(q_{\beta}) \nonumber \\
&-& V_{2 \rightarrow 1} (q_\gamma+q_\lambda \rightarrow q_\beta) \{\hat{a}^{\dagger}(q_{\gamma})\hat{a}^{\dagger}(q_{\lambda}) -\hat{a}^{\dagger}(q_{\beta})\} \frac{1}{2} \hat{a}(q_{\gamma})\hat{a}(q_{\lambda}
)]
\end{eqnarray}
\end{center}
Following the logic of Chapter  3.3  we consider a matrix element
\begin{center}
\begin{eqnarray}
\left\{\prod_{\alpha'}<\phi_{j+1}(q_{\alpha'})|\right\} (1+\mathcal{H}\Delta Y )\left\{|\prod_{\alpha}\phi_{j}(q_{\alpha})>\right\}
\end{eqnarray}
\end{center}
First, look at the second term of the Hamiltonian
\begin{center}
\begin{eqnarray}\label{appB_4}
\left\{\prod_{\alpha'}<\phi_{j+1}(q_{\alpha'})|\right\}
 \sum_{\beta} \sum_{\gamma} \sum_{\lambda} [-V_{1\rightarrow 2}
(q_\beta\rightarrow q_\gamma+q_\lambda)\hat{a}^{\dagger}(q_{\gamma})\hat{a}^{\dagger}(q_{\lambda})\hat{a}(q_{\beta})]
 \left\{\prod_{\alpha}|\phi_{j}(q_{\alpha})>\right\} \nonumber \\
\left\{\prod_{\alpha'}<\phi_{j+1}(q_{\alpha'})|\right\}
 \sum_{\beta} \sum_{\gamma} \sum_{\lambda} [-V_{1\rightarrow 2}
(q_\beta\rightarrow q_\gamma+q_\lambda)
\phi^*_{j+1}(q_{\gamma})
\phi^*_{j+1}(q_{\lambda})
\phi_{j}(q_{\beta})
]
 \left\{\prod_{\alpha}|\phi_{j}(q_{\alpha})>\right\}
\end{eqnarray}
\end{center}
In Eq.~(\ref{appB_4}) we used the property of the coherent states given by  Eq.~(\ref{appB_11}).

In the continuous limit $\delta_{\alpha,\beta}$ is replaced by $\delta(q_\alpha-q_\beta)$, and
$\prod_{\alpha}d\phi^*(q_\alpha)d\phi(q_\alpha)$  by  functional integration $\int \mathcal{D} \phi^* \mathcal{D} \phi$.

From here we see that rest of the calculations is similar to that of Chapter  3.3 and we end up with the expression
for a matrix element of an operator $A$ between states of initial $Y_0$ and final rapidity $Y$
\begin{center}
\begin{eqnarray}\label{appB_3}
<Y|A|Y_0>  \sim \int \mathcal{D}\Phi^+ \mathcal{D}\Phi A(Y) e^{S}
\end{eqnarray}
\end{center}
where
\begin{eqnarray}\label{appB_5}
 S\, &=& \,\int \left( \int\Phi^+(q)  \frac{d}{dY} \Phi(q)dq +\mathcal{H} ( 1-\Phi , -\Phi^+ ) \, \right) dY
\end{eqnarray}
with the Hamiltonian given by

\begin{eqnarray}
\mathcal{H} &=& \nonumber
\\
 &+&\int d^4 q_0 d^4 q_1 d^4 q_2 [V_{1\rightarrow 2}
(q_0\rightarrow q_1+q_2) \{-\Phi(q_{0}) +\Phi(q_{1})
+\Phi(q_{2})-\Phi(q_{1})\Phi(q_{2}) \} \Phi^+(q_{0}) \nonumber \\
&+& V_{2 \rightarrow 1} (q_1+q_2 \rightarrow q_0) \{-
\Phi(q_{1})\Phi(q_{2})+\Phi(q_{1}) +\Phi(q_{2})-\Phi(q_{0}) \}
\frac{1}{2} \Phi^+(q_{1})\Phi^+(q_{2})
 ]
\end{eqnarray}


\begin{thebibliography}{99}
\bibitem{GLR}
L. V. Gribov, E. M. Levin and M. G. Ryskin, {\it Phys. Rep.}\, 
{\bf 100}, 1 (1983).

\bibitem{MUQI}
A. H. Mueller and J. Qiu,  {\it Nucl. Phys.},427 {\bf B 268} 
(1986) .

\bibitem{BFKL}
 E. A. Kuraev, L. N. Lipatov, and F. S. Fadin, {\it  Sov. Phys. 
JETP}
                {\bf 45}, 199 (1977); \,\,\,
Ya. Ya. Balitsky and L. N. Lipatov,
               {\it   Sov. J. Nucl. Phys.}\, {\bf 28}, 22 (1978).

\bibitem{BART}
J.~Bartels, M.~Braun and G.~P.~Vacca,
 { \it Eur.\ Phys.\ J.}  {\bf C40}, 419 (2005)
  [arXiv:hep-ph/0412218]\,;\,\,\,
J.~Bartels and C.~Ewerz,
{\it JHEP} {\bf 9909}, 026 (1999)
[arXiv:hep-ph/9908454]\,;\,\,\,
J.~Bartels and M.~Wusthoff,
{\it Z.\ Phys.} {\bf C66}, 157 (1995)\,;\,\,\,
\,\,\,\,A.~H.~Mueller and B.~Patel,
{\it Nucl.\ Phys.}  {\bf B425}, 471 (1994)
[arXiv:hep-ph/9403256];\,\,\,
J.~Bartels,
Z.\ Phys.\  {\bf C60}, 471 (1993).

\bibitem{BRN}
M.~A.~Braun,
{\it   Phys.\ Lett.}\,  {\bf B632} (2006) 297
  [arXiv:hep-ph/0512057];\,\,
arXiv:hep-ph/0504002\,;
{ \it Eur.\ Phys.\ J.}  {\bf C16}, 337 (2000)
[arXiv:hep-ph/0001268];\,\,\,
  Phys.\ Lett.\ B {\bf 483} (2000) 115
  [arXiv:hep-ph/0003004];\,\,
  Eur.\ Phys.\ J.\ C {\bf 33} (2004) 113
  [arXiv:hep-ph/0309293];\,\,\,
{\it Eur.\ Phys.\ J.}  {\bf C6}, 321 (1999)
[arXiv:hep-ph/9706373];\,\,\,
M.~A.~Braun and G.~P.~Vacca,
{\it Eur.\ Phys.\ J.}  {\bf C6}, 147 (1999)
[arXiv:hep-ph/9711486].
\bibitem{NP}
H.~Navelet and R.~Peschanski,
{\it Nucl.\ Phys.}  {\bf B634}, 291 (2002)
[arXiv:hep-ph/0201285]; \,\,\,
{\it Phys.\ Rev.\ Lett.}  {\bf 82}, 137 (1999),
[arXiv:hep-ph/9809474];\,\,\,
{\it Nucl.\ Phys.}\  {\bf B507}, 353 (1997) 
[arXiv:hep-ph/9703238].



\bibitem{BLV}
J.~Bartels, L.~N.~Lipatov and G.~P.~Vacca,
{\it Nucl.\ Phys.}  {\bf B706}, 391 (2005)
[arXiv:hep-ph/0404110].
\bibitem{GRC}
V.~N.~Gribov,
{\it   Sov.\ Phys.\ JETP} {\bf 26}, 414 (1968)
  [{\it Zh.\ Eksp.\ Teor.\ Fiz.}  {\bf 53}, 654 (1967)].
\bibitem{KAN}
O.V. Kancheli, {\it Sov. JETP Lett.} {\bf 18}, 274  (1973),
in Caneschi, L. (ed.): ``Regge theory of low-p(T) hadronic interactions',  p. 312, North-Holland, 1989;
{\it Pisma Zh.\ Eksp.\ Teor.\ Fiz.}\,  {\bf 22} (1975) 491.

\bibitem{GRPO}
P.~Grassberger and K.~Sundermeyer, 
{\it Phys.\ Lett.} {\bf B77}, 220 (1978).
\bibitem{LEPO}
E.~Levin,
{\it Phys.\ Rev.} {\bf D49, 4469} (1994).
\bibitem{BOPO}
K.~G.~Boreskov,
{\it ``Probabilistic model of Reggeon field theory,''}
arXiv:hep-ph/0112325 and reference therein.

\bibitem{GARD}
C.W. Gardiner,{\it ``Handbook of Stochastic Methods for Physics, Chemistry
and the
Natural
Science"}, Springer-Verlag, Berlin, Heidelberg 1985.

\bibitem{VEN}
~G. ~Veneziano, {\it Phys. Letters} {\bf 52B} (1974) 220; {\it Nucl.
Phys.}{\bf B74} (1974) 365;\\
~M. ~Ciafaloni, ~G. ~Marchesini anf ~G. ~Veneziano,
{\it Nucl. Phys.} {\bf B98} (1975) 493.
\bibitem{IM}
E.~Iancu and A.~H.~Mueller,
{\it  Nucl.\ Phys.}\,  {\bf A730} (2004) 460, 494,
[arXiv:hep-ph/0308315],[arXiv:hep-ph/0309276].
\bibitem{KOLE}
M.~Kozlov and E.~Levin,
{\it Nucl.\ Phys.}\  {\bf A739} (2004) 291
[arXiv:hep-ph/0401118].


\bibitem{MUCD}
A.~H.~Mueller, {\it Nucl.\ Phys.} {\bf B415}, 373 (1994);
{\it ibid}  {\bf B437}, 107 (1995).

\bibitem{L1}
E.~Levin and M.~Lublinsky,
{\it   Nucl.\ Phys.}\, {\bf A730}, 191 (2004)
  [arXiv:hep-ph/0308279].
\bibitem{L2}
E.~Levin and M.~Lublinsky,
{\it   Phys.\ Lett.} {\bf B607}, 131 (2005)
  [arXiv:hep-ph/0411121].
\bibitem{L3}
E.~Levin and M.~Lublinsky,
{\it   Nucl.\ Phys.}  {\bf A763},172 (2005) ,
  arXiv:hep-ph/0501173.
\bibitem{L4}
 E.~Levin,
{\it   Nucl.\ Phys.}  {\bf A763}, 140 (2005),   arXiv:hep-ph/0502243.
\bibitem{IT}
E.~Iancu and D.~N.~Triantafyllopoulos,
{\it   Nucl.\ Phys.}\  {\bf A756}, 419 (2005)
  [arXiv:hep-ph/0411405];\,\,{\it Phys.\ Lett.} {\bf B610}, 253 (2005)
  [arXiv:hep-ph/0501193].
\bibitem{MSW}
 A.~H.~Mueller, A.~I.~Shoshi and S.~M.~H.~Wong,
{\it   Nucl.\ Phys.} {\bf B715}, 440 (2005)
  [arXiv:hep-ph/0501088].


\bibitem{MUSH}
A.~H.~Mueller and A.~I.~Shoshi,
  {\it ``Small-x physics near the saturation regime,''}
  arXiv:hep-ph/0405205;\,\,\,{\it Nucl.\ Phys.} \, {\bf B692} (2004) 175
  [arXiv:hep-ph/0402193].
\bibitem{KKM}
G.~P.~Korchemsky, J.~Kotanski and A.~N.~Manashov,
Phys.\ Lett.\  {\bf B583} (2004) 121
[arXiv:hep-ph/0306250] \,\,\,;
S.~E.~Derkachov, G.~P.~Korchemsky, J.~Kotanski and A.~N.~Manashov,
Nucl.\ Phys.\  {\bf B645} (2002) 237
[arXiv:hep-th/0204124];\,\,\,
S.~E.~Derkachov, G.~P.~Korchemsky and A.~N.~Manashov,
Nucl.\ Phys.\  {\bf B661} (2003) 533
[arXiv:hep-th/0212169];\,\,\,
Nucl.\ Phys.\  {\bf B617} (2001) 375
[arXiv:hep-th/0107193];\,\,\,
and references therein.


\bibitem{LI}
L.~N.~Lipatov,
{\it Phys.\ Rept.}  {\bf 286}, 131 (1997)
[arXiv:hep-ph/9610276];\,\,Sov. Phys. JETP {\bf 63}, 904 (1986) and
references therein.
\bibitem{IST}
E.~Iancu, G.~Soyez and D.~N.~Triantafyllopoulos,
  arXiv:hep-ph/0510094.


\bibitem{B}
I.~Balitsky,
[arXiv:hep-ph/9509348];\,\,
{\it Phys.\ Rev.} {\bf D60}, 014020 (1999)
[arXiv:hep-ph/9812311].




\bibitem{K}
Y.~V.~Kovchegov,
{\it Phys.\ Rev.}  {\bf D60}, 034008  (1999),
[arXiv:hep-ph/9901281].
\bibitem{KOVG}
Y.~V.~Kovchegov,
  Phys.\ Rev.\ D {\bf 72} (2005) 094009
  [arXiv:hep-ph/0508276].
\bibitem{LELE}
L. ~Pechenik and ~H.~Levine, {\it Phys. Rev.} \,{\bf E59}, 3893
(1999),arXiv:cond-mat/9811020.


\bibitem{HH}
H. Hinrichsen, {\it Advances in Physics} {\bf  49} 815 (2000). 


	


\bibitem{WE}
H.~Weigert,
 {\it Prog.\ Part.\ Nucl.\ Phys.}\,  {\bf 55}, 461 (2005),
arXiv:hep-ph/0501087 and references therein.


\bibitem{BIW}
J.~P.~Blaizot, E.~Iancu and H.~Weigert,
{\it Nucl.\ Phys.}\,  {\bf A713}, 441 (2003),
[arXiv:hep-ph/0206279].

\bibitem{DMS}
C.R. Doering, C. Mueller and P. Smereka,{\it  Physica}\, {\bf A 325}, 243 (2002).
\bibitem{MSX} 
C.~Marquet, G.~Soyez and B.~W.~Xiao,
   {\it ``On The Probability Distribution Of The Stochastic Saturation Scale In
  QCD,''}
  arXiv:hep-ph/0606233.
\bibitem{SO}
 G.~Soyez,
 {\it  Phys.\ Rev.}\  {\bf D72}, 016007 (2005),
  [arXiv:hep-ph/0504129].


 
\bibitem{JIMWLK}
~J.~Jalilian-Marian, A.~Kovner, A.~Leonidov and H.~Weigert,
{\it  Phys.\ Rev.}\,  {\bf D59}, 014014 (1999),
[arXiv:hep-ph/9706377];\,\,  {\it Nucl.\ Phys.}\,{\bf B504}, 415  
(1997),
[arXiv:hep-ph/9701284]; \,\,\,
J.~Jalilian-Marian, A.~Kovner and H.~Weigert,
  {\it Phys.\ Rev.}  {\bf D59}, 014015 (1999),
  [arXiv:hep-ph/9709432];\,\,\,
 A.~Kovner, J.~G.~Milhano and H.~Weigert,
 {\it  Phys.\ Rev.}  {\bf D62}, 114005 (2000),
  [arXiv:hep-ph/0004014]\,; \,\,\,
E.~Iancu, A.~Leonidov and L.~D.~McLerran,
{\it  Phys.\ Lett.}\,  {\bf B510}, 133 (2001);
[arXiv:hep-ph/0102009];\,\, {\it  Nucl.\ Phys.}\,  {\bf A692}, 583 
(2001),
[arXiv:hep-ph/0011241];\,\,\,
E.~Ferreiro, E.~Iancu, A.~Leonidov and L.~McLerran,
 {\it  Nucl.\ Phys.}\  {\bf A703}, 489 (2002),
  [arXiv:hep-ph/0109115];\,\,\,
H.~Weigert,
{\it  Nucl.\ Phys.}  {\bf A703}, 823 (2002),
[arXiv:hep-ph/0004044].
\bibitem{CION}
 M.~Ciafaloni and E.~Onofri,
 {\it  Nucl.\ Phys.}\,  {\bf B151}, 118 (1979).

\bibitem{BLAGK}
J.~Bartels and M.~G.~Ryskin,
  Z.\ Phys.\  {\bf C76} (1997) 241
  [arXiv:hep-ph/9612226].




\bibitem{MV}
L. McLerran and R. Venugopalan, {\it  Phys. Rev.}  {\bf D 49},2233,
3352  (1994); {\bf D 50},2225 (1994); {\bf D 53},458 (1996); {\bf
D 59},09400
(1999).
\bibitem{AGK}
V.~A.~Abramovsky, V.~N.~Gribov and O.~V.~Kancheli,
  { \it Yad.\ Fiz.}\,  {\bf 18},  595 (1973)
  [{\it Sov.\ J.\ Nucl.\ Phys.}\,  {\bf 18}, 308 (1974)].

\bibitem{GRIB1}
 V.~N.~Gribov,
  {\it ``Space-Time Description Of Hadron Interactions At High Energies,''}
  arXiv:hep-ph/0006158.
\bibitem{LEPAGE}
 G.~P.~Lepage and S.~J.~Brodsky,
  Phys.\ Rev.\ D {\bf 22} (1980) 2157.
\bibitem{FEYN}
 R.~P.~Feynman,
  Phys.\ Rev.\ Lett.\  {\bf 23} (1969) 1415.
\bibitem{KOLU}
A.~Kovner and M.~Lublinsky,
ArXiv:hep-ph/0510047;\,\,
arXiv:hep-ph/0503155;\,\,
{\it Phys.\ Rev.\ Lett.}\,  {\bf 94}, 181603 (2005)
  [arXiv:hep-ph/0502119];\,\,
{\it JHEP} {\bf 0503}, 001 (2005)
  [arXiv:hep-ph/0502071];\,\,
{\it Phys.\ Rev.}\,  {\bf D71}, 085004 (2005)
  [arXiv:hep-ph/0501198];\,\,
  {\it ``Odderon and seven Pomerons: QCD Reggeon field theory from JIMWLK
  evolution,''}
  arXiv:hep-ph/0512316.
\bibitem{HIMS}
Y.~Hatta, E.~Iancu, L.~McLerran and A.~Stasto,
{\it Nucl.\ Phys.}\, {\bf A762} (2005) 272
  [arXiv:hep-ph/0505235];
 Nucl.\ Phys.\ A {\bf 764} (2006) 423
  [arXiv:hep-ph/0504182].
\bibitem{LIEF}
~L.~N. ~Lipatov, {\it Nucl. Phys.} {\bf B452}, 69 (1995).

\bibitem{RY}
I. Gradstein and I. Ryzhik, {\it `` Tables of Series, Products, and
Integrals"}, Verlag MIR, Moskau,1981.

\bibitem{KAMKE}
 Erich Kamke,  {\it ``Differentialgleichungen: Lösungsmethoden  und Lösunge
  Bd. 1: Gewöhnliche
Differentialgleichungen,"} Stuttgart, Germany,
  Teubner Verlag, 1983; \\
Andrey D. Polyanin, {\it ``Handbook of Linear Differential Equations for Engineers and Scientists"}, 
Chapman \& Hall/CRC, 2002. 
\bibitem{LT}
E.~Levin and K.~Tuchin,
{\it Nucl.\ Phys.}\  {\bf A693} (2001) 787
[arXiv:hep-ph/0101275];\,\,\,{\bf A691} (2001) 779
[arXiv:hep-ph/0012167];\,\,\,{\bf B573} (2000) 833
[arXiv:hep-ph/9908317];\,\,\,\,\,M.~Kozlov and E.~Levin,
{\it Nucl.\ Phys.}\  {\bf A764} (2001) 498
  arXiv:hep-ph/0504146.
\bibitem{IIM}
E.~Iancu, K.~Itakura and L.~McLerran,
 {\it  Nucl.\ Phys.}\,	  {\bf A708} (2002) 327
  [arXiv:hep-ph/0203137].

\bibitem{THEORVRSDATA}
N.~Armesto and M.~A.~Braun,
  {\it Eur.\ Phys.\ J.}\  {\bf C20}, 517 (2001)
  [arXiv:hep-ph/0104038];\,\,
M.~Lublinsky,
  {\it Eur.\ Phys.\ J.}\  {\bf C21}, 513 (2001)
  [arXiv:hep-ph/0106112];\,\,\,\,
E.~Levin and M.~Lublinsky,
 {\it   Nucl.\ Phys.}  {\bf A712}, 95 (2002)
  [arXiv:hep-ph/0207374];\,\,
  {\it Nucl.\ Phys.}  {\bf A712}, 95 (2002)
  [arXiv:hep-ph/0207374];\,\,
{\it Eur.\ Phys.\ J.}\, {\bf C22}, 647 (2002)
  [arXiv:hep-ph/0108239];\,\,\,\,
M.~Lublinsky, E.~Gotsman, E.~Levin and U.~Maor,
 {\it   Nucl.\ Phys.}\,  {\bf A696}, 851 (2001)
  [arXiv:hep-ph/0102321];\,\,
 {\it  Eur.\ Phys.\ J.}\,  {\bf C27}, 411 (2003)
  [arXiv:hep-ph/0209074];\,\,\,\,
K.~Golec-Biernat, L.~Motyka and A.Stasto,
 {\it  Phys.\ Rev.} {\bf D65}, 074037 (2002)
  [arXiv:hep-ph/0110325];\,\,\,
E.~Iancu, K.~Itakura and S.~Munier, {\it 
  Phys.\ Lett.}\, {\bf B590} (2004) 199
  [arXiv:hep-ph/0310338].
K.~Rummukainen and H.~Weigert,
{\it   Nucl.\ Phys.} \, {\bf A739}, 183 (2004)
  [arXiv:hep-ph/0309306];\,\,K.~Golec-Biernat and A.~M.~Stasto,
 {\it  Nucl.\ Phys.} {\bf B668}, 345 (2003)
  [arXiv:hep-ph/0306279];\,\,\,\,E.~Gotsman, M.~Kozlov, E.~Levin, U.~Maor and E.~Naftali,
 {\it   Nucl.\ Phys.}\, {\bf A742}, 55 (2004)
  [arXiv:hep-ph/0401021];\,\,\,\,K.~Kutak and A.~M.~Stasto,
 {\it  Eur.\ Phys.\ J.}\,  {\bf C41}, 343 (2005)
  [arXiv:hep-ph/0408117];\,\,\,\,G.~Chachamis, M.~Lublinsky and A.~Sabio Vera,
{\it   Nucl.\ Phys.}  {\bf A748}, 649 (2005)
  [arXiv:hep-ph/0408333];\,\,\,\,
  J.~L.~Albacete, N.~Armesto, J.~G.~Milhano, C.~A.~Salgado and U.~A.~Wiedemann,
 {\it  Phys.\ Rev.} {\bf D71}, 014003 (2005)
  [arXiv:hep-ph/0408216];\,\,\,\,E.~Gotsman, E.~Levin, U.~Maor and E.~Naftali,
{\it   Nucl.\ Phys.} {\bf A750} (2005) 391
  [arXiv:hep-ph/0411242].


\bibitem{MS}
A.~H.~Mueller and G.~P.~Salam,
{\it  Nucl.\ Phys.}\,  {\bf B475}, 293 (1996),
[arXiv:hep-ph/9605302];\,\,\,\,G.~P.~Salam,
{\it   Nucl.\ Phys.}\,  {\bf B461}, 512 (1996).

\bibitem{BON}
  S.~Bondarenko,
  {\it ``Langevin equation in effective theory of interacting QCD pomerons in the
  limit of large  $N_c$ ''}
  arXiv:hep-ph/0703045.
\end{thebibliography}
\end{document}